\documentclass[11pt]{article}
\usepackage{fullpage}
\usepackage{authblk}
\usepackage{setspace}
\usepackage{geometry}
\usepackage[utf8]{inputenc}
\usepackage[T1]{fontenc}

\DeclareUnicodeCharacter{03BC}{\ensuremath{\mu}}
\usepackage{amsmath}
\usepackage{amssymb}
\usepackage{graphicx}
\usepackage{color}
\usepackage{subfig}
\usepackage{bm}
\usepackage{gensymb}
\usepackage{bigints}
\usepackage[usenames,dvipsnames,svgnames,table]{xcolor}
\usepackage{pifont}

\usepackage{orcidlink}
\usepackage{physics}
\usepackage{makeidx}

\usepackage[sorting=none]{biblatex}
\usepackage{epsfig}
\usepackage{colortbl}
\usepackage{colordvi}
\usepackage{verbatim}
\usepackage{pdfpages}
\usepackage[absolute,overlay]{textpos}
\usepackage{ragged2e}
\usepackage{caption}
\usepackage{hyperref}
\hypersetup{
    colorlinks=true,
    linkcolor=blue,
    filecolor=Violet,     
    urlcolor=RubineRed,
    citecolor=red
}

\begin{document}
\date{}
\author{Ritwik Acharyya \thanks{\href{mailto:ritwikacharyya@kgpian.iitkgp.ac.in}{ritwikacharyya@kgpian.iitkgp.ac.in}}}
\author{Pritam Banerjee \thanks{\href{mailto:pritam@phy.iitkgp.ac.in}{pritam@phy.iitkgp.ac.in}}}
\author{Sayan Kar \thanks{\href{mailto:sayan@phy.iitkgp.ac.in}{sayan@phy.iitkgp.ac.in}}}
\affil[]{Department of Physics, Indian Institute of Technology Kharagpur, 721302, India}
\title{Modelling Einstein cluster using Einasto profile} 

\maketitle

\begin{abstract}
We demonstrate a general relativistic approach to model dark matter halos using the Einstein cluster, with the matter stress-energy generated by collisionless particles moving on circular geodesics in all possible angular directions and orbital radii. Such matter, as is known, allows an anisotropic pressure profile with non-zero tangential but zero radial pressure. We use the Einasto density profile for the Einstein cluster. Analytical studies on its properties (metric functions) and stability issues are investigated. Further, to establish this model (with the Einasto profile) as one for a dark matter halo,
we use the SPARC galactic rotation curve data and estimate the best-fit values for the model parameters. General relativistic features (beyond the Keplerian velocities) such as the tangential pressure profile, are quantitatively explored. Thus, Einstein clusters with the Einasto profile,  which tally well with observations, may be considered as a viable model for dark matter halos. 

\end{abstract}

\section{Introduction}
In 1939, Einstein \cite{einstein1938gravitational} presented a model known as the Einstein cluster, consisting of a large number of non-interacting particles that are moving along circular geodesics under their collective gravitational field. The particles in the cluster can move either in the same orbit with different phases or in different orbits with various orbital radii and orientations. Consequently, the cluster with a large number of particles becomes a continuous yet random distribution of collisionless particles characterized by the density or the total angular momentum \cite{geralico2012einstein} profile. It is a static and spherically symmetric exact solution of Einstein's equation (provided a density profile is chosen), which is centrifugally stable, i.e., the outward centrifugal force on a particle counterbalances the inward-directed gravity. It turns out that such an Einstein cluster has an anisotropic pressure profile with zero radial and non-zero tangential pressure. Being tangential pressure dominated, Einstein clusters are studied in the literature as an example of an anisotropic energy-momentum tensor. Einstein clusters can be used in modelling a globular cluster of stars \cite{Bohmer_2007} as they provide a good approximation to the average density of the matter profile and the average value of the gravitational field. Also, this model can describe any galactic rotation curve (galactic rotation curve is the mean circular velocity around the center of the galaxy as a function of galactocentric distance measured in the disk midplane \cite{Ablimit2020}) by specifying its anisotropy. Due to this feature, Einstein clusters can indeed be considered as a viable model for a dark matter halo \cite{lake2006galactic}.  

Einstein cluster models have been extensively studied in the literature by Zapolsky \cite{1968ApJ...153L.163Z}, Hogan \cite{Hogan1978}, Florides \cite{44e0e859-ce8d-33fc-9e45-490604ed12e4}, and Comer \cite{G_L_Comer_1993}. In this paper, we have considered Einstein's static model, but the natural generalization of a dynamical model was investigated by Datta \cite{Datta1970} and Bondi \cite{Bondi1971} without specifying explicit metric functions. In recent years, the gravitational collapse scenario has been studied explicitly. In this context, Mahajan et al. \cite{Mahajan_2007} have explicitly presented the critical collapse scenario for Einstein cluster models. Magli \cite{Giulio_Magli_1997} presented a general class of time-dependent spherically symmetric solutions of Einstein's equations with zero radial pressure, and later on, Gair \cite{J_R_Gair_2001} discussed all possible kinds of evolution of the shells constituting the cluster. Einstein clusters have also been studied as a toy model for small-scale inhomogeneous spacetimes by Szybka and Rutkowski \cite{Szybka_2020}. Zero radial pressure allows the construction of a static spacetime with small-scale inhomogeneities with a physical equation of state.

In our work, we will use the Einstein cluster with a chosen density profile as a model for dark matter halos. The characterization of dark matter halos is usually done by means of their spherically averaged density profiles. Several heuristic fitting functions have been proposed to capture the shapes of individual dark matter halos, e.g., Navarro–Frenk–White (NFW) profile \cite{1997ApJ...490..493N}, Moore profile \cite{Moore:1999gc}, DC14 profile \cite{Cintio_2014}, Burkert type density profile \cite{burkert1995structure}), and luminous matter distributions, e.g.,  de-Vaucouleurs profile \cite{1948AnAp...11..247D}, the Jaffe profile \cite{10.1093/mnras/202.4.995}. Though the insight for such density profiles arise from analytic studies, the exact mathematical form emerges from cosmological N-Body simulations \cite{1997ApJ...490..493N}. Without a proper understanding of how these density profiles can be derived from the more fundamental principles, it is an open question why these fits appear to tally well with the simulated and observed density profiles. There are few works in the literature where these density profiles have been derived using a statistical approach (e.g., see \cite{Wagner:2020opz}).


Recent N-body CDM simulations suggest \cite{2004MNRAS.349.1039N} that Einasto density profile \cite{1965TrAlm...5...87E, 1969Afz.....5..137E} provides an excellent fit to a wide range of dark matter halos. In this model, the logarithmic density slope shows a power law behaviour rather than the density itself. This model was proposed in \cite{2004MNRAS.349.1039N} and \cite{Merritt_2005}, where it was argued to be a better fit than the NFW model \cite{Retana_Montenegro_2012}. The Einasto model has been extensively used to describe the simulated dark matter halos \cite{Merritt_2006} and to fit the observed galactic rotation curve (e.g., \cite{Chemin_2011,Li_2020,refId1,refId2}). However, to the best of our knowledge, there has been no work to incorporate the Einasto profile in an Einstein cluster, possibly due to the fact that it is difficult to handle analytically and demands a numerical treatment. In this paper,
we use the Spitzer Photometry and Accurate Rotation Curve (SPARC) \cite{Lelli_2016} database to show the Einstein cluster with the Einasto profile as a feasible dark matter halo model. From the best fit of the rotation curve for 175 late-type galaxies, we estimate the value of the free parameters present in the Einasto profile. The tangential pressure profile has also been shown in this model as the signature of general relativistic effects.  

This paper is organized as follows. In section 2, the theoretical framework of the Einstein cluster is reviewed, and a detailed calculation of rotational velocity is shown. In section 3, the Einasto profile is introduced, and the analytic properties of the profile required for our analysis are discussed. In section 4, the stability analysis of the cluster and its classification based on stability issues are shown. In section 5, the SPARC database is used to find the best fit Einstein cluster model using the Einasto profile.  

\section{Theoretical construction of Einstein cluster}
\subsection{Stress-energy inside an Einstein cluster}
Let us consider a static, spherically symmetric distribution of particles moving along circular geodesics about the center of symmetry. The associated line element can be written as 
\begin{equation}\label{eqn:metric}
    ds^2 = -e^{\alpha} c^2 dt^2+e^{\beta}dr^2+r^2\left(d\theta^2+\sin^2\theta d\phi^2\right).
\end{equation}
Here $\alpha$ and $\beta$ are the functions which depend only on the radial coordinates. Let us consider the orthonormal tetrad for a static observer,
\begin{equation}
    e_{t}= e^{-\alpha/2}c^{-1}\partial_t , \hspace{0.1in} e_{r}=e^{-\beta/2}\partial_r , \hspace{0.1in}  e_{\theta}=\frac{1}{r}\partial_{\theta}, \hspace{0.1in}  e_{\phi}=\frac{1}{r\sin\theta}\partial_{\phi}. \label{tetrad}
\end{equation}
Now, according to the ansatz given by Einstein \cite{einstein1938gravitational}, the average stress-energy tensor can be taken as,
 \begin{equation}
    \langle T^{\mu \nu}\rangle=\frac{n}{m}\langle p^{\mu}p^{\nu}\rangle,
    \label{eqn:Tmunu}
\end{equation}
where $n$ is the proper number density of the particles having rest mass $m$. $p^\mu=m u^\mu$ is the particle 4-momentum where $u^\mu$ is the 4-velocity satisfying the geodesic equation. The averaging $\langle...\rangle$ is done over all trajectories with all the directions and phases passing through a spatial point where the energy-momentum tensor of the orbiting particles is computed. The above ansatz has an underlying condition that $n(r)$ at any layer is independent of other layers. This holds trivially true for circular trajectories. For a general case of a thick Einstein shell constructed using elliptical orbits that pass through the outer surface but not necessarily through the inner surface, the above ansatz may not hold (see \cite{GL}). 

We can write the 4-velocity associated with the spherical orbits in the static observer's frame as,
\begin{equation}
 u=\gamma[c e_{t}+v^{\theta}e_{\theta}+v^{\phi}e_{\phi}],\hspace{0.3in}
    \gamma=\left[1-(v^{\theta}/c)^2 - (v^{\phi}/c)^2\right]^{-1/2}.
    \label{eqn:U}
\end{equation}
There are two constants of motion associated with each trajectory, 
\begin{equation}
    p_{\mu}\xi_{(t)}^{\mu} =-m c^2 \gamma e^{\alpha/2}=-E, \hspace{0.3in}  p_{\mu}\xi_{(\phi)}^{\mu} =m\gamma v^{\phi}r\sin\theta=L_z.
\end{equation}
Here, $\xi^\mu_{(t)} = c e^{\alpha/2} e_{t} $ and $\xi^\mu_{(\phi)} = r \sin\theta e_{\phi}$ are two Killing vectors expressed in the observer's basis. $E$ and $L_z$ are the energy and z component of the angular momentum of a particle. For an equatorial orbit ($\theta =\pi/2$), total angular momentum $L = L_z$, thus, $ (v^{\phi})^2 = L^2/m^2 \gamma^2 r^2 $. For any off-equatorial circular orbit, $L$ has contributions from both $v^{\theta}$ and $v^{\phi}$. Therefore, taking the average over all types of trajectories, we can write,
\begin{equation}
    \langle (v^{\theta})^2 \rangle = \langle (v^{\phi})^2 \rangle = \frac{L^2}{2 m^2 \langle \gamma^2 \rangle r^2}, \hspace{0.1in} \text{where} \hspace{0.3in} \langle \gamma^2 \rangle = 1 + \frac{L^2}{m^2 c^2 r^2}.
    \label{eqn:v}
\end{equation}
Now, using Eq. (\ref{eqn:Tmunu}), (\ref{eqn:U}) and (\ref{eqn:v}), we can write, 
\begin{align}
     \langle T^t_{~t} \rangle &= -\langle T^{tt} \rangle = - n m c^2\left(\frac{m^2 c^2 r^2 +L^2}{m^2 c^2 r^2}\right) \equiv -\rho c^2, \nonumber \\
     \langle T^r_{~r} \rangle &= \langle T^{rr} \rangle = 0, \nonumber \\
    \langle T^\theta_{~\theta} \rangle &= \langle T^{\theta\theta} \rangle = \langle T^\phi_{~\phi} \rangle = \langle T^{\phi\phi} \rangle = - n m \left(\frac{L^2}{2 m^2 r^2}\right) \equiv P_t, 
\end{align}
where $P_t$ is the tangential pressure and $\rho c^2$ is the energy density. All the other components of $T^\mu_{~\nu}$ are zero. At this point, we mention the approach taken by \cite{GL}, where the authors have considered a thick Einstein shell consisting of elliptical orbits whose turning points pass through the upper and lower surfaces. They have demonstrated that $\langle T^r_{~r} \rangle = 0 $ is the only valid option leading to circular geodesics.

\subsection{Metric functions within the Einstein cluster}
Now, as we have obtained the stress-energy tensor of the system, we can use the Einstein equation to relate $\rho c^2$ and $P_t$ with the metric functions.  
The relevant Einstein equations corresponding to the metric in Eq. (\ref{eqn:metric}) are, 
\begin{align}
  & \frac{1}{r^2}[r(1-e^{-\beta})]^\prime =8\pi \frac{G}{c^4}\rho c^2, \label{eqn:Gtt}\\ &
   \alpha^\prime=\frac{1}{r}(e^\beta -1), \label{eqn:Grr}\\ &
   \frac{e^{-\beta}}{2}\left[ \alpha^{\prime\prime}+\frac{\alpha^{\prime 2}}{2}+\frac{\alpha^\prime -\beta^\prime}{r}-\frac{\beta^\prime \alpha^\prime}{2}\right]=8\pi \frac{G}{c^4} P_t. \label{eqn:Gphiphi}
\end{align}
Since radial pressure is zero for this model, from Eq. (\ref{eqn:Grr}), we can relate the metric components. Solving Eq. (\ref{eqn:Grr}) algebraically for $\beta$ we get,
\begin{equation}
    \beta=\ln (1+r\alpha^\prime).
\label{eqn:beta}
\end{equation}
Substituting Eq. (\ref{eqn:Gtt}) and (\ref{eqn:Grr}) in Eq. (\ref{eqn:Gphiphi}) we get a relation between $P_t$ and $\rho c^2$,
\begin{equation}
    P_t= \frac{\rho c^2}{4}\left(e^\beta-1\right)=\frac{r\alpha^\prime}{4}\rho c^2. \label{eqn:p_t}
\end{equation}
It can be shown that both angular momentum distribution ($L$) and total number distribution ($mn$) can be written explicitly in terms of the metric functions \cite{geralico2012einstein},
\begin{align}
    L &=\frac{m r c\sqrt{r\alpha^\prime/2}}{\sqrt{1-r\alpha^\prime/2}} \label{eqn:Ltilde},\\
    mn&= \frac{c^2}{4\pi G r}\frac{1-r \alpha^\prime/2}{(1+r \alpha^\prime)^2}\left[r \alpha^{\prime\prime}/2 + r (\alpha^\prime)^2 /2 + \alpha^\prime \right]=(1-r\alpha^\prime/2)\rho.\label{eqn:mn}
\end{align}
As $\alpha$ can be directly computed from observations of galactic rotation curves \cite{PhysRevLett.92.051101}, one always has the freedom to alter the angular momentum distribution and their number distribution ($mn$) to match the observed rotation curve exactly \cite{lake2006galactic}. 

If we choose the density profile of the cluster $\rho(r)$, we obtain the gravitational mass $M(r)$, which can be used to compute the metric functions. The gravitational mass within a spherical region of radius $r$ is given by \cite{PhysRev.136.B571},
\begin{equation}
    M(r)=4\pi\int_0^r \rho r^2 dr.
\label{eqn:spherical mass distribution}
\end{equation}
Now, integrating Eq. (\ref{eqn:Gtt}) and using Eq. (\ref{eqn:Grr}) along with Eq. (\ref{eqn:spherical mass distribution}), we get the metric functions as,
\begin{equation}
    e^\alpha =\left[1-\frac{2GM/c^2}{R}\right]e^{-2\phi(r)},~~~ e^\beta = \left[1-\frac{2GM(r)/c^2}{r}\right]^{-1},
    \label{eqn:gtt}
\end{equation}
where,
\begin{equation}
    \phi (r)=\int_r^R \frac{GM(r)/c^2}{r\left(r-2GM(r)/c^2\right)} dr.
    \label{eqn:phi}
\end{equation}
$M$ is the total mass of the cluster.
Although the definition of $M(r)$ in Eq. (\ref{eqn:spherical mass distribution}) resembles the Newtonian mass of a spherical distribution, it turns out that the gravitational mass explicitly appears in the metric function $e^\beta$ via the Einstein equation in Eq. (\ref{eqn:Gtt}). One can also follow the opposite route by choosing the form of $e^\beta$ and obtain Eq. (\ref{eqn:spherical mass distribution}).  

\subsection{Rotational velocity inside the cluster}
Corresponding to the metric given by Eq. (\ref{eqn:metric}), timelike circular geodesics conform to two constants of motion. Considering the equatorial plane without any loss of generality, they are $\Tilde{E} = c^2 e^\alpha \Dot{t}$, and $ \Tilde{L} = r^2 \Dot{\phi} $. Using the normalization of 4-velocity, we can write,
\begin{equation}
    \Dot{r}^2 + V_{\text{eff}}(r) = \Tilde{E}^2,~~~ \text{where} ~~~ 
    V_{\text{eff}}=c^2e^{\alpha/2}\sqrt{1+\frac{\Tilde{L}^2}{r^2 c^2}}. \label{eqn:Veff}
\end{equation}
For a circular orbit to exist, $\Dot{r} = 0 $, which means
\begin{equation}
  -\frac{E^2}{\Tilde{L}^2}+\frac{c^2 e^\alpha}{r^2}=-\frac{c^4e^\alpha}{\Tilde{L}^2}. \label{eqn:EandtildeL}
\end{equation}
In addition, we also have $\Ddot{r}=0$, which yields $V_{\text{eff}}^\prime (r)=0$, i.e.,
\begin{equation}
    \Tilde{L}=\frac{rc \sqrt{r\alpha^\prime/2}}{\sqrt{1-r\alpha^\prime/2}} =  \frac{rc\sqrt{GM(r)/c^2}}{\sqrt{r-3GM(r)/c^2}}. \label{eqn:Ltildemass}
\end{equation}
The above expression is exactly the same as Eq. (\ref{eqn:Ltilde}), whereas, for the second part, we have used Eq. (\ref{eqn:gtt}).

Now, the observed rotational velocity is measured by a far away coordinate observer as $ v_r = r\frac{d\phi}{dt}$. Using Eq. (\ref{eqn:EandtildeL}), and Eq. (\ref{eqn:Ltildemass}) we get 
\begin{equation}
    v_r = r\frac{d\phi}{dt}= r \frac{\Dot{\phi}}{\Dot{t}}=\sqrt{\frac{e^\alpha}{\left(r^2/\Tilde{L}^2 +1/c^2\right)}} = c\sqrt{e^\alpha\frac{GM(r)/c^2}{r-2GM(r)/c^2}}. \label{eqn:rotvel}
\end{equation} 

\section{Properties of Einasto profile}

Einasto density profile, first presented by Einasto for modelling the galactic halos of M31, M32, M87, and the Milky Way \cite{1965TrAlm...5...87E},  is characterized by a power-law logarithmic slope \cite{Retana_Montenegro_2012},
\begin{equation}
    \gamma(r)\equiv -\frac{d\ln \rho (r)}{d\ln r} \propto r^{1/n}, \label{eqn:logarithmic slope}
\end{equation}
where $n$ is the Einasto index, a positive number that defines the steepness of the power law. The density profile can be found by integrating Eq. (\ref{eqn:logarithmic slope}),
\begin{equation}
    \rho (r)=\rho_s \exp\left\{-d_n\left[\left(\frac{r}{r_s}\right)^{1/n}-1 \right]\right\}, \label{eqn:Einasto profile1}
\end{equation}
where $r_s$ is the radius of the sphere containing half of the total mass, $\rho_s$ is the mass density at $r=r_s$, and $d_n$ is a numerical constant that constrains $r_s$ as the half-mass radius.
As $r_s$ encloses half of the total mass, $d_n$ can be found from the solution of the equation 
$2\Gamma (3n,d_n)=\Gamma(3n)$ \cite{Retana_Montenegro_2012, 2022A&A...667A..47B}.\\
From Eq. (\ref{eqn:logarithmic slope}), an equivalent form of this density profile can be written, which has been used for calculational purposes, 
\begin{equation}
  \rho (r)=\rho_0 \exp\left[- 
  \left(\frac{r}{h}\right)^{1/n}\right]. \label{eqn:Einasto profile2}
\end{equation}
Here $\rho_0$ is the central density and $h$ is the scale length defined by,
$\rho_0=\rho_s e^{d_n}$ and  $h=r_s/d_n^n$ respectively.  \\
Now, the total mass of an Einstein cluster with an Einasto profile is,
\begin{equation}
    M=\int_0^\infty \rho (r) 4\pi r^2 dr = 4\pi \rho_0 h^3 n \Gamma (3n). \label{eqn:Einasto mass}
\end{equation}
The central density can be written in terms of the total mass given by Eq.  (\ref{eqn:Einasto mass}). Substituting in Eq. (\ref{eqn:Einasto profile2}) and redefining, $ s=r/h $ we get, 
\begin{equation}
    \rho (r)=\frac{M}{4\pi h^3 n \Gamma (3n)} e^{-s^{1/n}}. \label{eqn:density in terms of total mass}
\end{equation}
At small radii, the density profile can be expanded and expressed as \cite{Retana_Montenegro_2012}
\begin{equation}
    \rho (s)= \frac{M}{4\pi h^3 n \Gamma (3n)} \left(1-s^{1/n}+...\right). \label{eqn:rho expansion}
\end{equation}
However, no physical density profile can be extended up to infinity. Hence, we cut off the density profile at $r=R$ and will assume the radius of the cluster to be $r=R$. The total mass of the cluster will be $M(R)=M$ where the mass function is defined by,
\begin{align}
    M(r) = \int_0^r \rho(r) 4\pi r^2 dr &= \int_0^\infty \rho (r) 4\pi r^2 dr-\int_r^\infty \rho (r) 4\pi r^2 dr \nonumber \\
    &=4\pi \rho_0 h^3 n\left[\Gamma(3n)-\Gamma\left(3n, \left(\frac{r}{h}\right)^{1/n}\right)\right], \label{eqn: Mass profile(r)}
\end{align}
where, the incomplete gamma function is defined as, $\Gamma (\alpha,x)=\int_x^\infty t^{\alpha -1}e^{-t}dt $.
Therefore, we can rewrite the mass profile as,
\begin{equation}
    M(r)=M\frac{\left[\Gamma(3n)-\Gamma\left(3n, \left(\frac{r}{h}\right)^{1/n}\right)\right]}{\left[\Gamma(3n)-\Gamma\left(3n, \left(\frac{R}{h}\right)^{1/n}\right)\right]}. \label{eqn: M(r)}
\end{equation}
As the mass function involves the gamma function and the incomplete gamma function, it would be very difficult to analyze this density profile analytically. We, therefore, use numerical methods to calculate the metric functions, and hence, we calculate the rotational velocities and compare them with the observed rotation curves. In Fig. (\ref{fig:Einasto_metric}), we plot the Einasto density (Eq. (\ref{eqn:Einasto profile2})), total angular momentum (Eq. (\ref{eqn:Ltilde})), and the metric functions (Eq. (\ref{eqn:gtt})) inside a Einstein cluster of radius $R=50M$. The density profile gets flatter for higher values of $h$ and $ n $. The bottom panels of the figure show the numerically computed metric functions where the interior metric ($r<R$) is continuously matched with the Schwarzschild metric (outside the cluster for $r>R$) at the boundary.
\begin{figure}[h]
\centering
\includegraphics[width=0.45\textwidth]{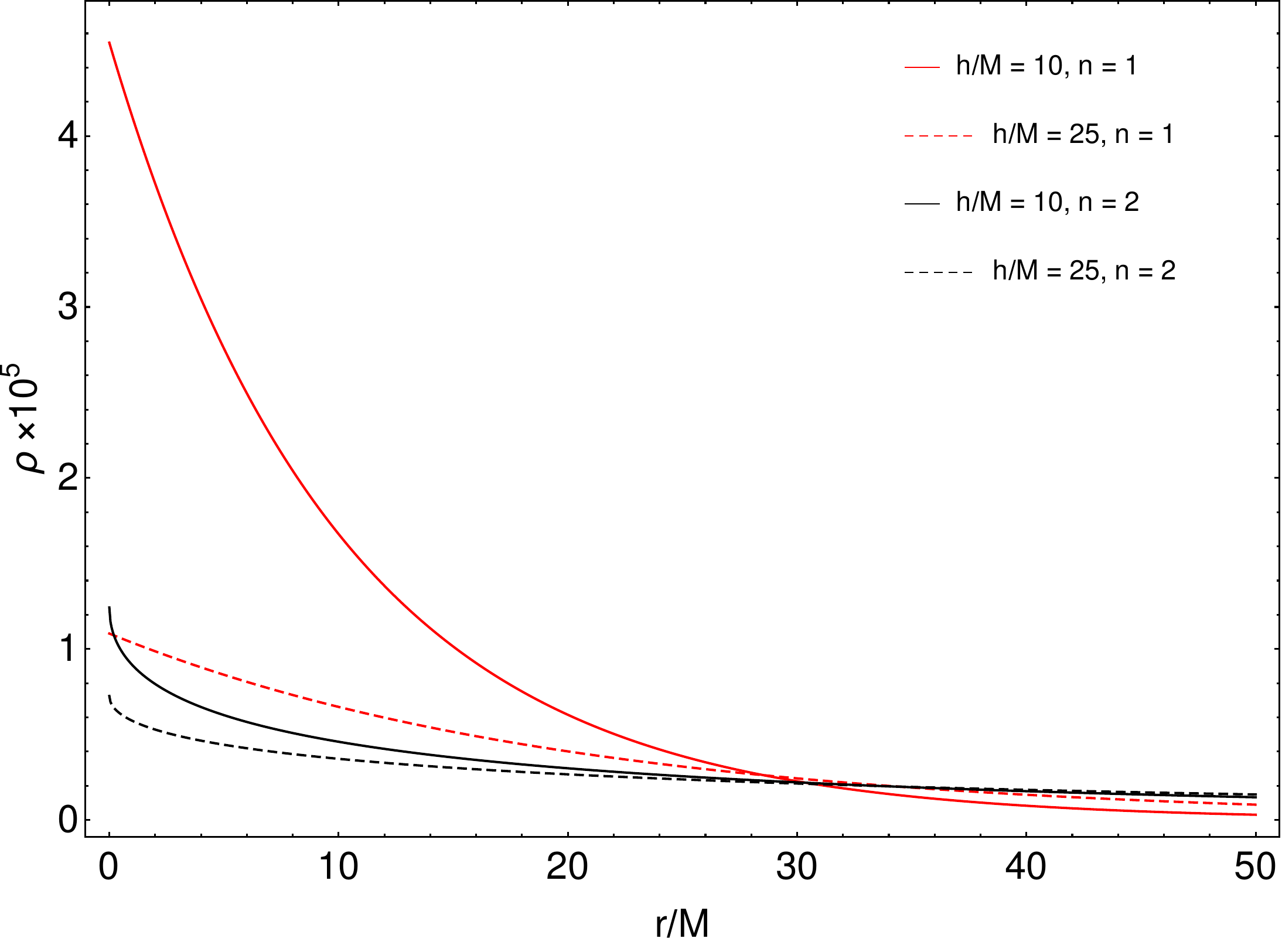}
\includegraphics[width=0.45\textwidth]{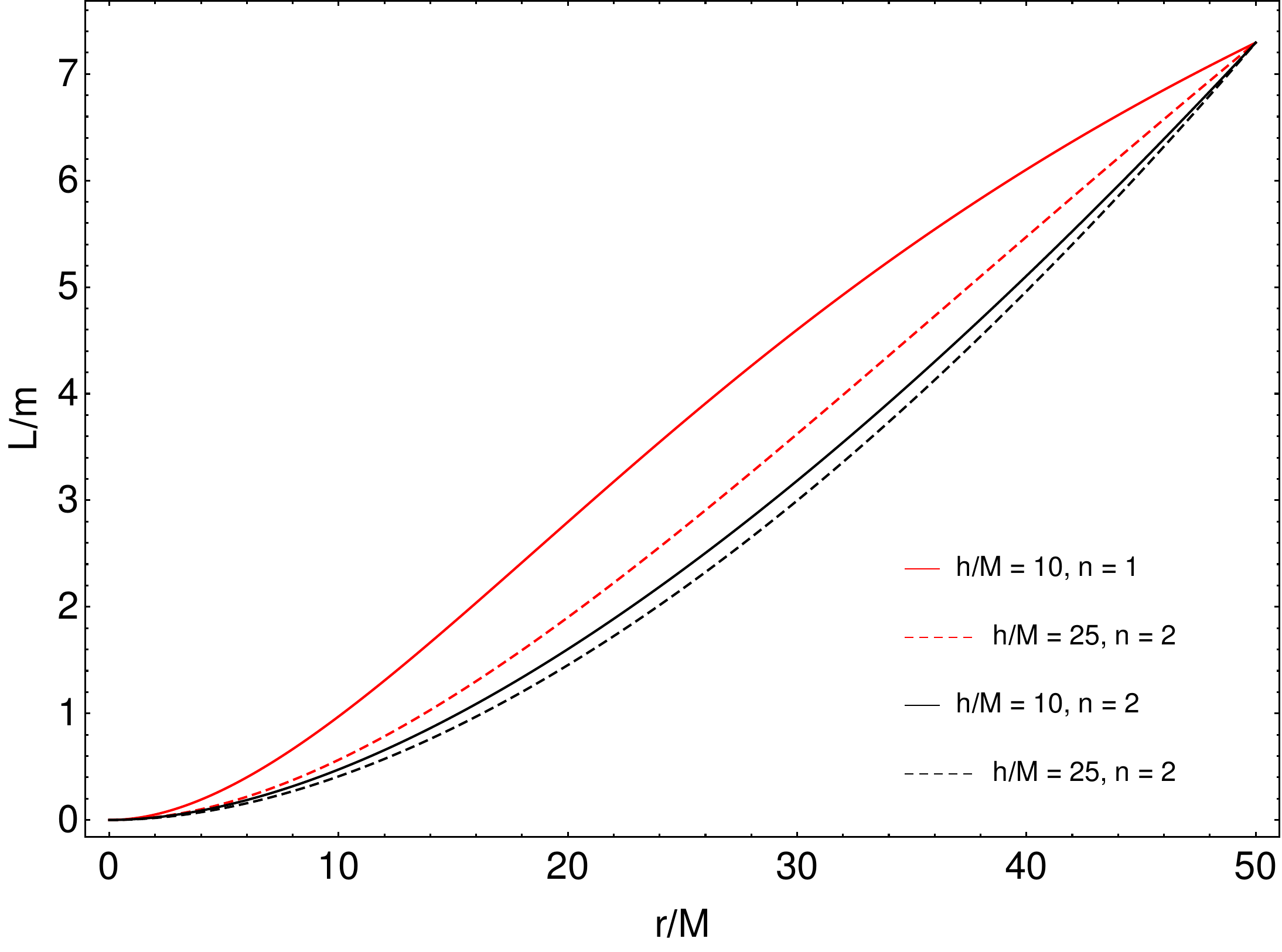}
\includegraphics[width=0.45\textwidth]{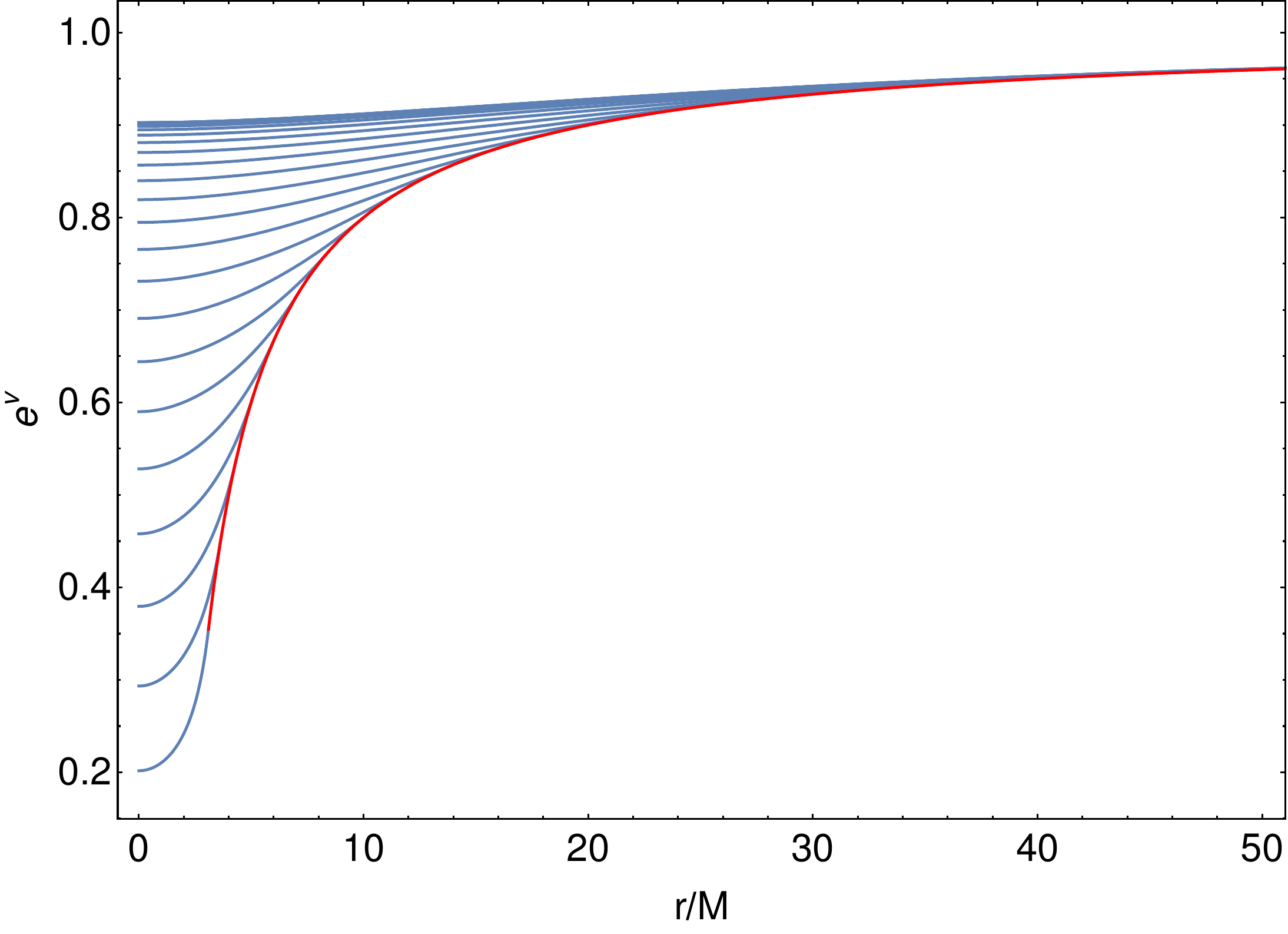}
\includegraphics[width=0.45\textwidth]{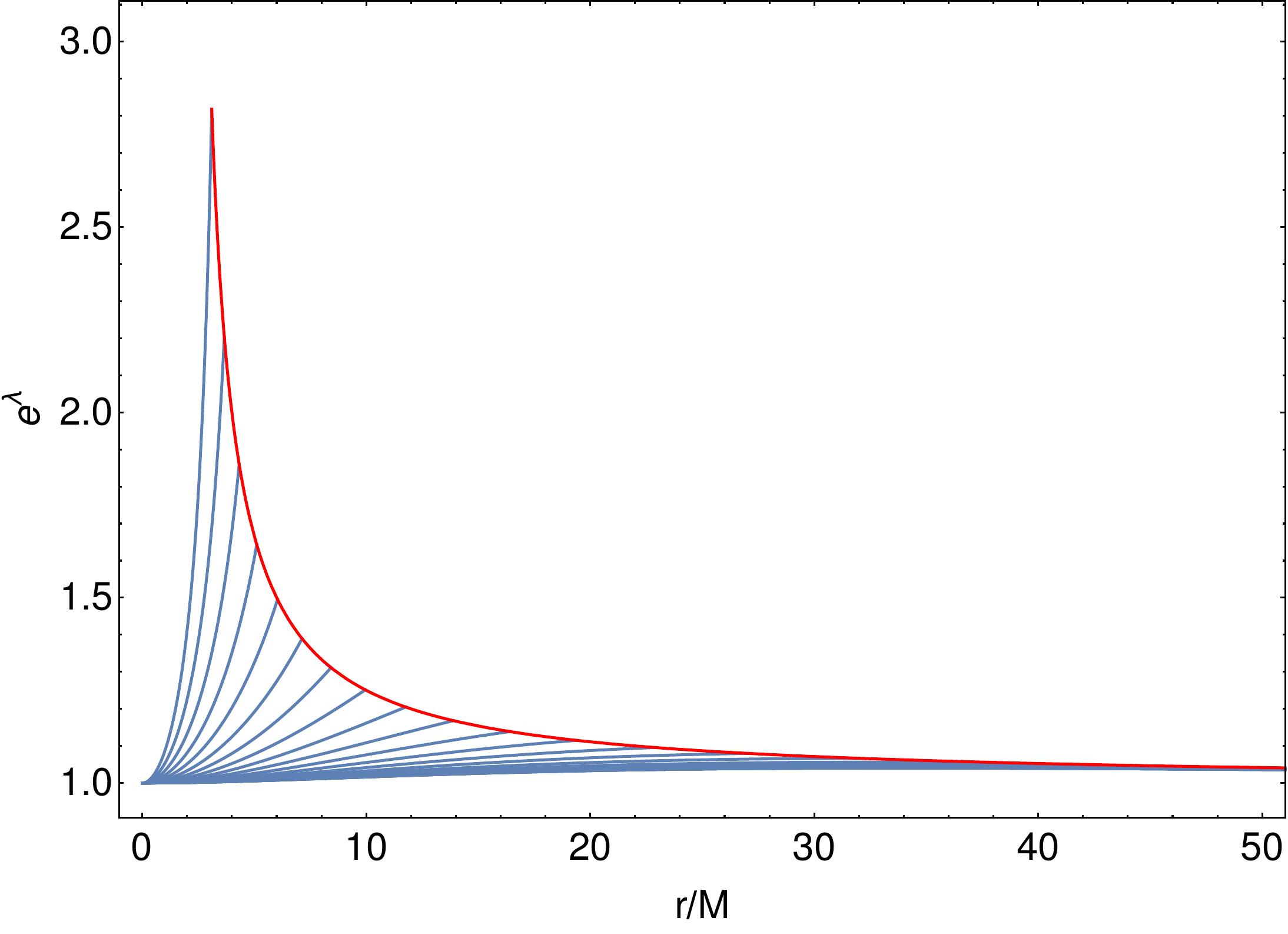}

\caption{\textbf{Top left panel:} Energy density of the Einasto profile is shown for an Einstein cluster of radius $R =50 M $. \textbf{Top right panel:} Total angular momentum distribution is depicted inside the cluster. \textbf{Bottom panels:} Metric components are plotted for the Einasto density profile (blue curves) inside the cluster radius $R$. Multiple curves shown in the figure stand for various values of $R$. At the surface of the cluster, blue curves, i.e., the metric functions inside the cluster, merge with the Schwarzschild metric functions (denoted by the red curves). For these bottom panels, we have chosen $h/M = 10$, $n=1$. \label{fig:Einasto_metric}}
\end{figure}

\section{Stability analysis of the cluster}
\subsection{Orbital stability inside the cluster}
For circular orbit to exist at any $r$, the effective potential in Eq. (\ref{eqn:Veff}) must have a minimum. By using $\Tilde{L}$ from Eq. (\ref{eqn:Ltilde}) in the condition $V_{\text{eff}}^\prime(r)=0$ we get, 
\begin{equation}
    0<r\alpha^\prime/2<1,\label{eqn:27}
\end{equation}
which is equivalent to the condition,
\begin{equation}
     r>3GM(r)/c^2. \label{eqn:28}
\end{equation}
The condition $V_{\text{eff}}^{\prime\prime}(r)>0$ will ensure the stability. Further differentiating $V_{\text{eff}}^\prime (r)$ we get, 
\begin{equation}
    r\alpha^{\prime\prime}-r(\alpha^\prime)^2+3\alpha^\prime >0, \label{eqn:Vprime}
\end{equation}
which is equivalent to
\begin{equation}
    \frac{d(\ln M(r))}{d(\ln r)}+1 -\frac{6M(r)}{r}>0. \label{eqn:massstability}
\end{equation}
When $M(r)=M$ is constant, this implies $r>6GM/c^2$, which is well known for the Schwarzschild metric.

As the gravitational field of the outermost layer of the system is given by the exterior Schwarzschild metric (boundary matching condition), the boundary radius of an Einstein cluster can never be less than $3M$. 
Based on the analysis of stability, the Einstein cluster can be classified into three different kinds of models \cite{1997NCimB.112..271C}:\\
Stable models constitute particles which move on stable circular orbits and for which the condition in Eq. (\ref{eqn:massstability}) is satisfied for every value of $r$ whether it is inside or outside the configuration. This implies that for every allowed value of $r$, the circular orbit corresponds to an absolute minimum of the effective potential. These kinds of clusters must obey the condition $R \geq 6M$ and are stable against radial perturbations. See the bottom panels of Fig. (\ref{fig:Einasto_metric}) where the effective potential conforms to a stable minimum everywhere inside the cluster. \\
Metastable models are those for which the condition in Eq. (\ref{eqn:massstability}) is satisfied in every internal point but not in an external region adjacent to the cluster. The perturbation of a metastable model produces a disappearance of the particles belonging to the outermost layers because orbits with $r\simeq R$ become unstable if perturbed to slightly larger $r$ \cite{geralico2012einstein}. See the top panels in Fig. (\ref{fig:Einasto_metric}), which show stable minima well inside the cluster. However, near the surface $V_{\text{eff}}(r) > V_{\text{eff}}(\infty) = 1 $ and there exist saddle points that are not stable against radial perturbations.    
Unstable clusters are configured by the particles that move on unstable circular orbits and do not satisfy the condition in Eq. (\ref{eqn:massstability}). 
 
\begin{figure}[h]
\centering
\includegraphics[width=0.45\textwidth]{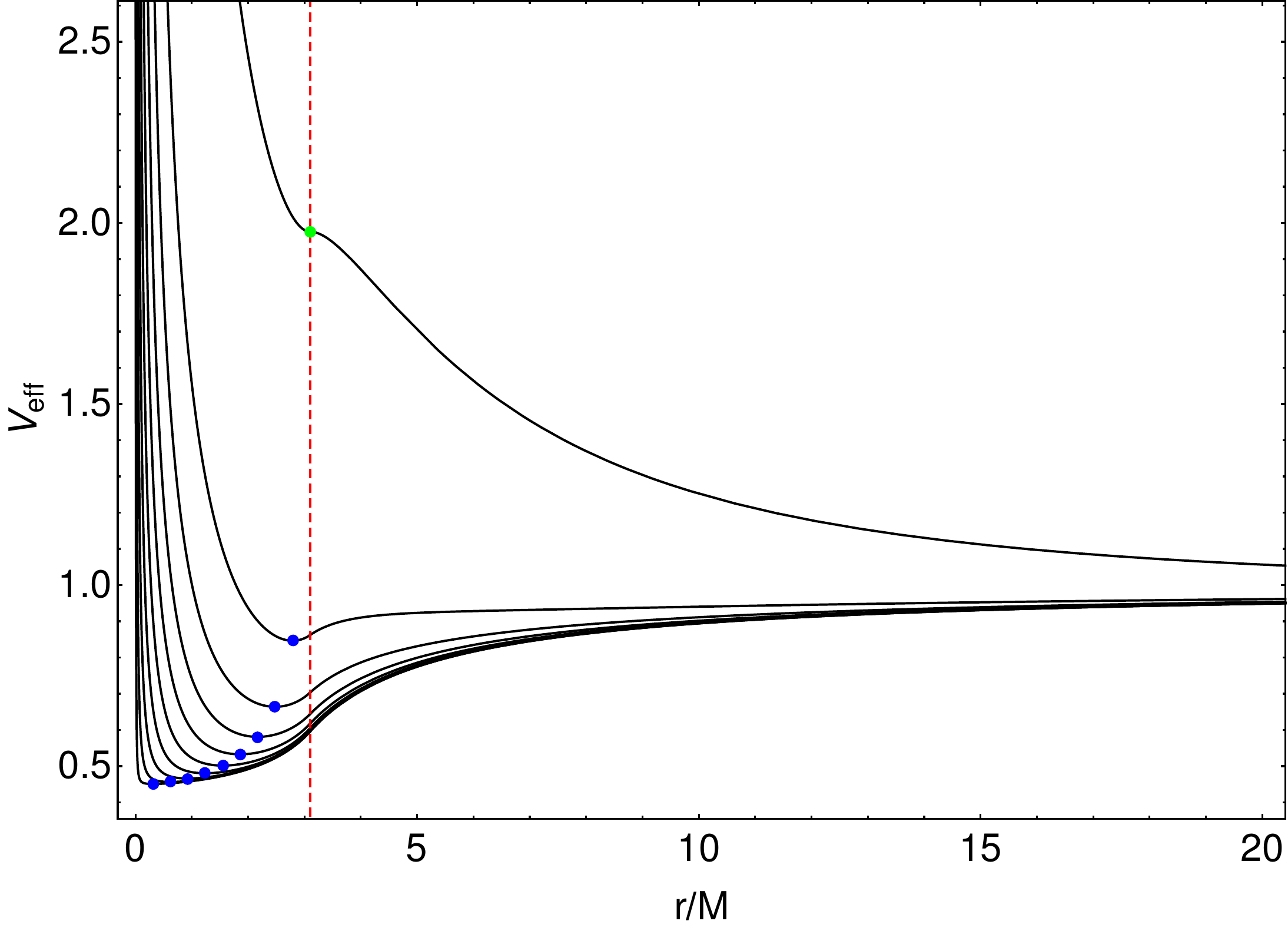}
\includegraphics[width=0.45\textwidth]{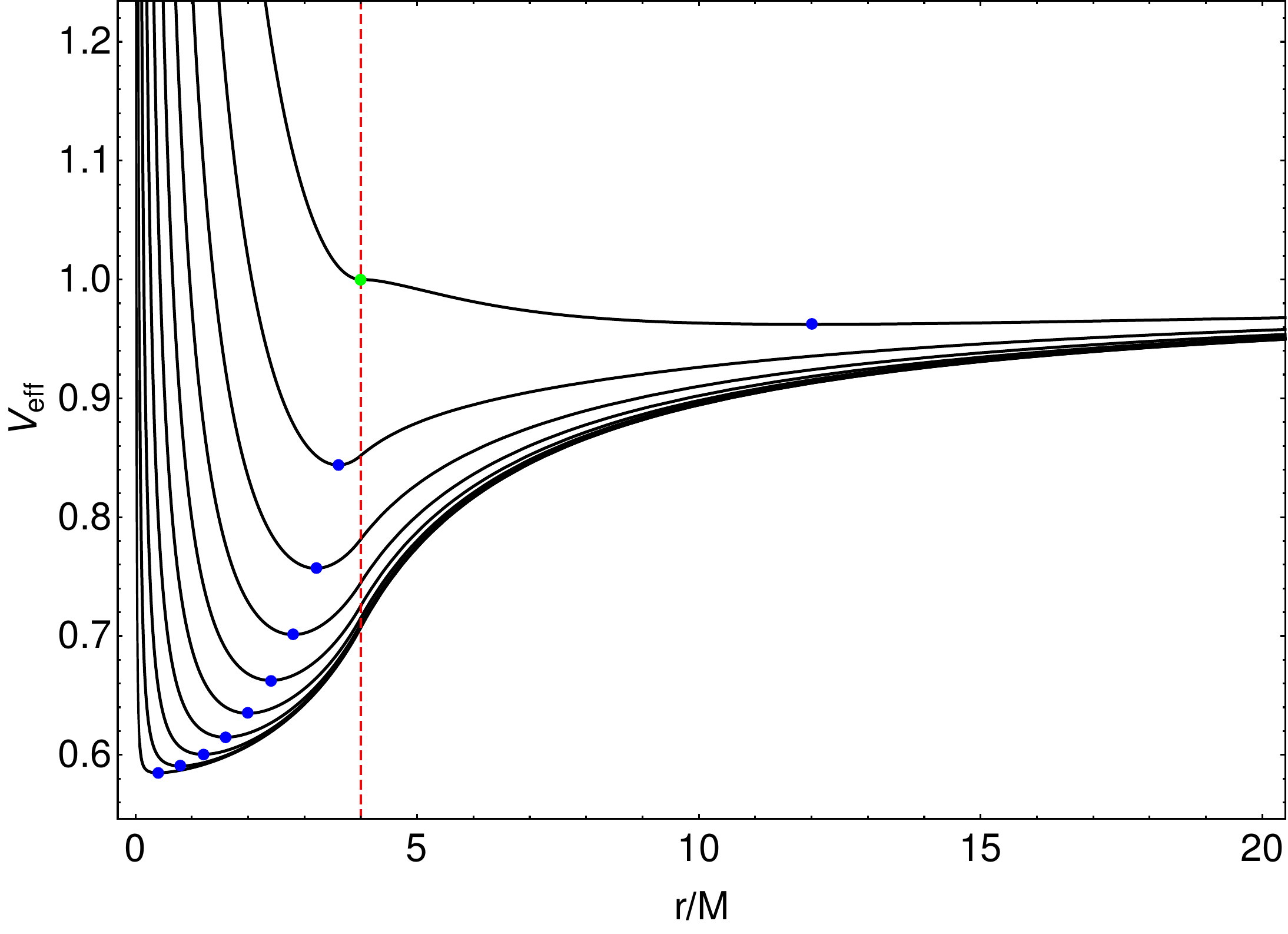}
\includegraphics[width=0.45\textwidth]{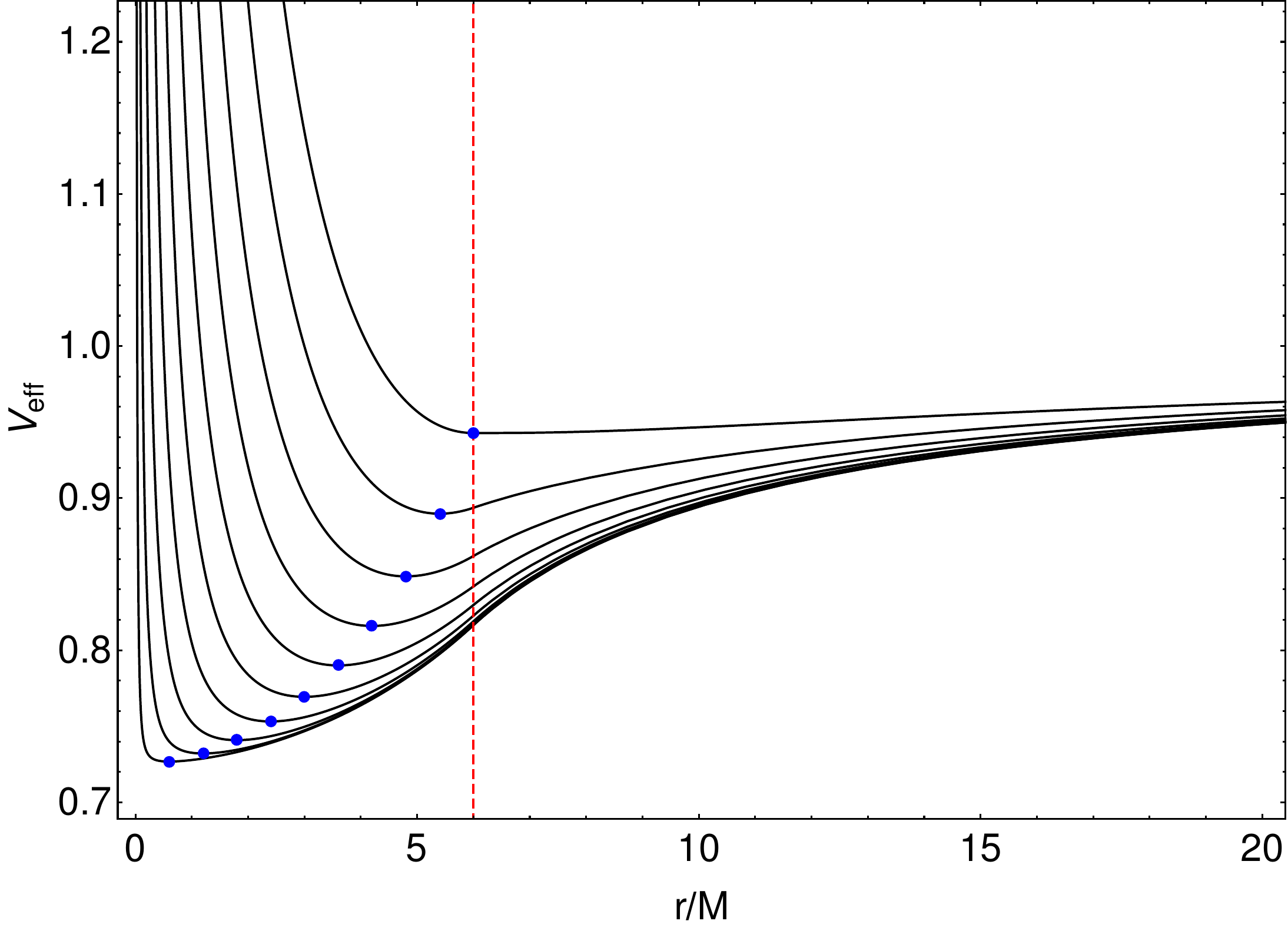}
\includegraphics[width=0.45\textwidth]{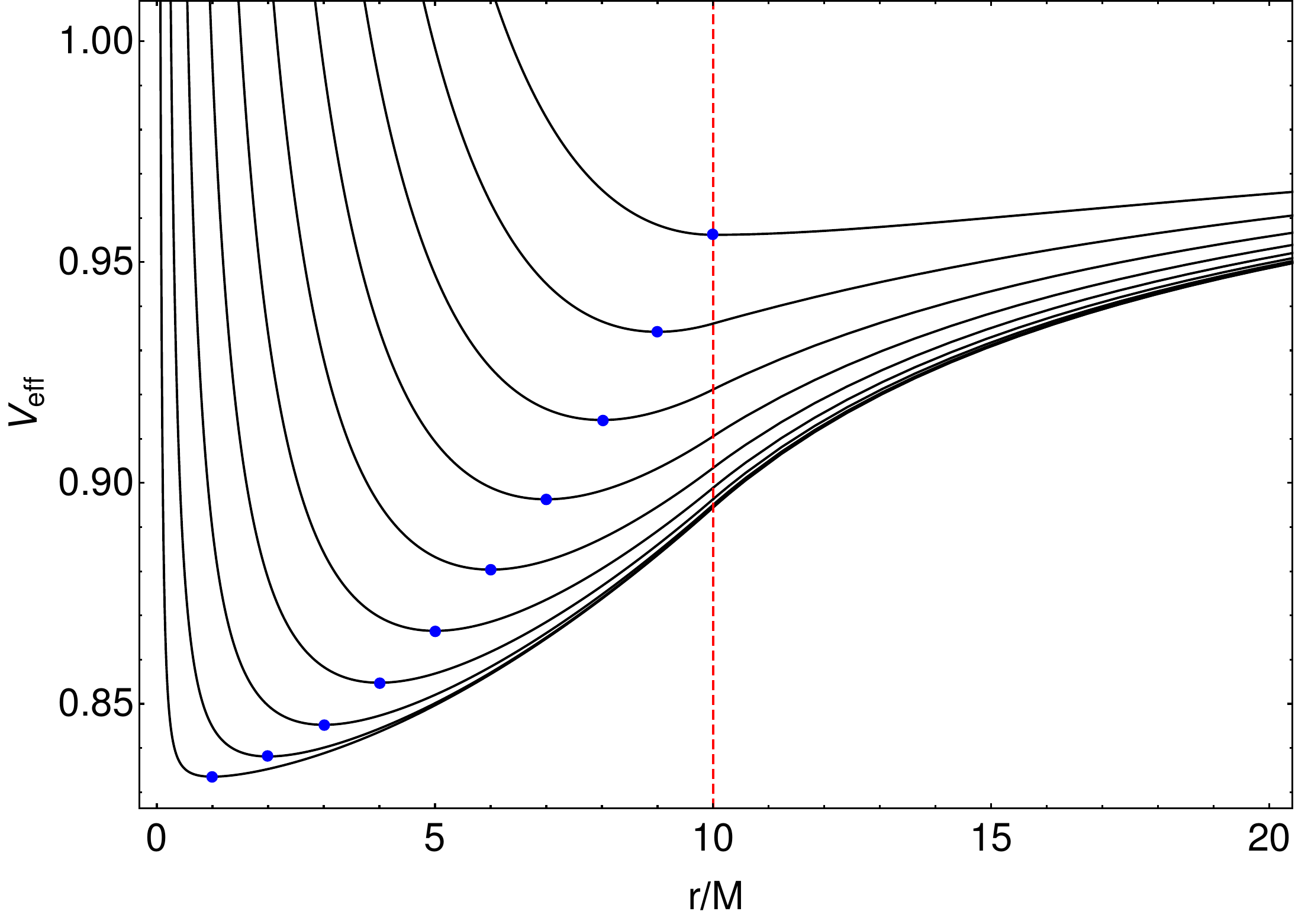}

\caption{Effective potential $V_\text{eff}$ is plotted for various cluster radii (red dashed vertical line) $R/M = 3.1$ (top left panel), $4$ (top right), $6$ (bottom left), and $10$ (bottom right). Multiple curves in each panel represent various values of $\Tilde{L}$. For this figure, we have chosen $h/M = 10$, $n = 1$. Blue dots indicate the minimum of the effective potential, where a stable circular trajectory is possible. For the top panels, we see a saddle point (unstable against perturbation) exactly at the surface, denoted as green dots. In the bottom left panel, the cluster is just stable, where the surface is made of innermost stable circular trajectories. The bottom right panel indicates that all the trajectories inside the cluster are stable. \label{fig:Einasto_metric}}
\end{figure}

\subsection{Binding energy analysis of the cluster}
Let us now turn to another method to analyse the stability of an Einstein cluster, 
adopted by Zapolsky \cite{1968ApJ...153L.163Z}. This is based on the behavior of the binding energy as the function of the cluster radius. The binding energy of the cluster is defined by 
\begin{equation}
    E_b=\frac{\textit{N}m-M}{\textit{N}m}, \label{eqn:fractional binding energy}
\end{equation}
where $\textit{N}m$ is the rest mass of the constituting particles, and $M$ is the total gravitational mass of the cluster (see Eq. \ref{eqn:spherical mass distribution}). It is similar to the fractional binding energy discussed in the context of nuclear physics. This binding energy represents the fraction of the total mass that turns into binding energy while the cluster is contracted from an infinite radius to the radius $R$. Thus, for stability, $E_b > 0$ should be satisfied.

\begin{figure}[]
\centering
\includegraphics[width=0.45\textwidth]{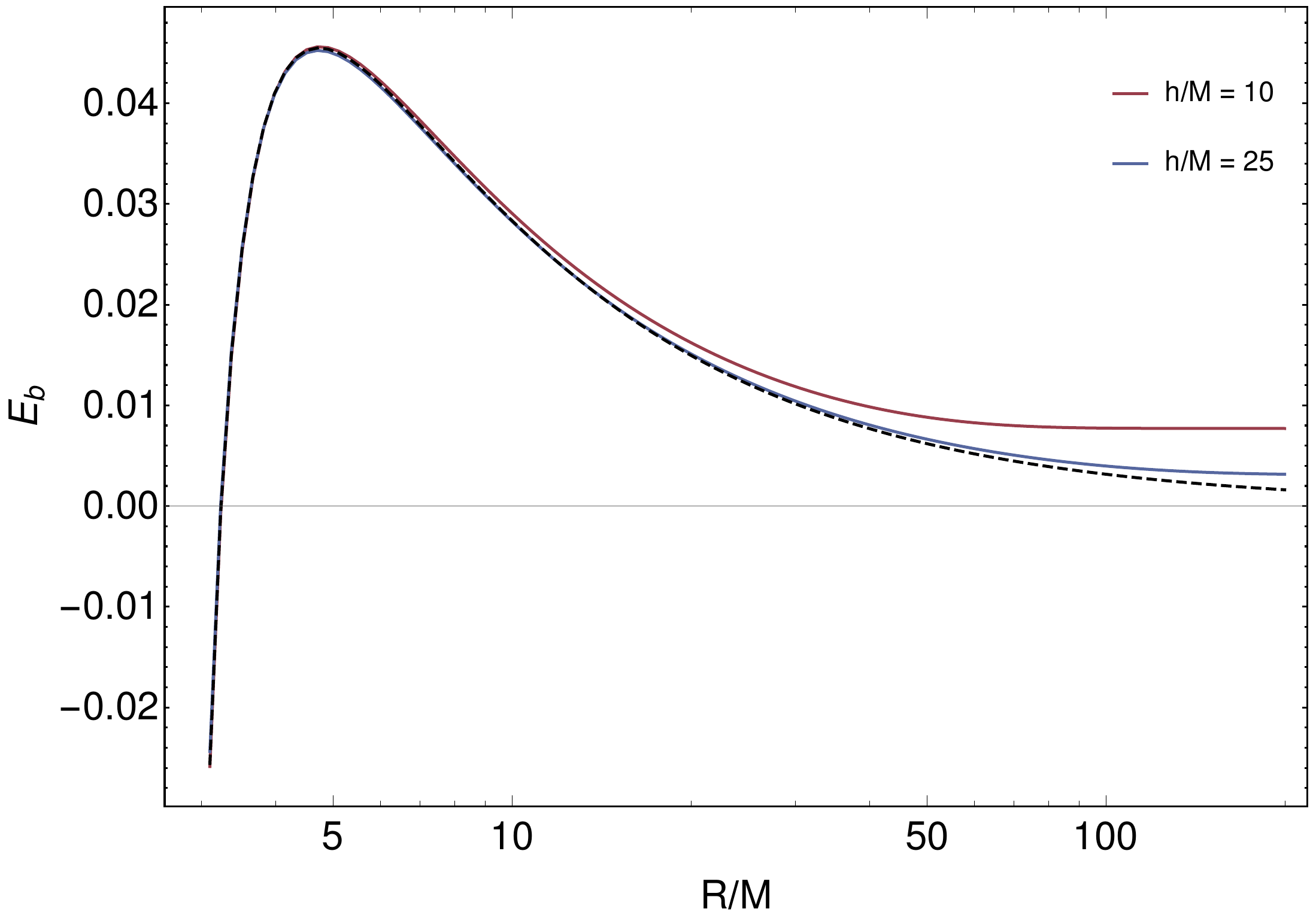}

\caption{Binding energy is plotted against $R/M$ for $n = 1$ and two different values of $h/M$ (shown in red and blue). Both curves visually superimpose for $n = 3$, as shown by the black dashed curve. }
\label{fig:Eb}
\end{figure}

From Eq. (\ref{eqn:mn}), we know the proper number density $n(r)$ in the rest frame of the particle. Now, in the static observer's frame, we know that the number density is $\gamma n(r)$ (due to special relativity). Since the total number of particles in the cluster is conserved, we can add up all the particles by considering static observers at each and every point inside the cluster. In order to do that, we first need to multiply the number density with the volume element in the observer's frame, given by $e^{\beta/2} r^2 \sin\theta dr d\theta d\phi$. Thus, the total number of particles is,
\begin{align}
    \textit{N}&= \int_0^R n(r)\gamma e^{\beta/2}r^2 dr \int_0^\pi \sin \theta d\theta \int_0^{2\pi} d\phi \nonumber \\
    &= 4\pi\int_0^R  n(r)\gamma e^{\beta/2}r^2 dr. \label{total rest mass}
\end{align}
In Fig. (\ref{fig:Eb}), the binding energy is plotted with $R/M$ for the Einstein cluster characterized by the Einasto profile. Two different values of $n$ ($n=1$ and $n=3$) and two different values of $h/M$ ($h/M=10$, $h/M=25$) are used. This curve remains positive for large values of $R/M$, representing stability. Now, as we decrease the value of $R/M$ gradually, $E_b$ increases and exhibits a maximum, indicating the most stable configuration. From Fig. (\ref{fig:Eb}), we find that the maximum of the effective binding energy is around $R=5M$. Below that, the cluster drastically changes its nature towards being unstable.  

\section{Estimation of the model parameters using SPARC data} 

In this section, our purpose is to use the Einasto profile to fit the Einstein cluster model with the observed galactic rotation curve data. We use the SPARC database \cite{Lelli_2016}, which contains the rotation curve data of 175 late-type galaxies having a wide range of surface brightness and luminosity obtained from the HI/H$\alpha$ measurements of the velocity profile of the galactic gas and the Spitzer photometry at $3.6 \mu m$. SPARC helps break the disc-halo degeneracy \cite{vanAlbada_1985} by accurately extracting the baryonic contribution in the rotation curve from the dark matter contribution using the fact that the stellar mass-to-light ratio is almost constant at the low energy infrared wavelength \cite{McGaugh_2014, Meidt_2014}. Thus, the database is used to verify various DM halo profiles. Here, we intend to model the halo using the proposed Einstein cluster with an Einasto profile and estimate the model parameters using the data. In the database, the mass-to-light ratios corresponding to the galactic bulge and the disk are set as one for convenience. Therefore, we also need to find their proper scaling during the fitting.  

The galactic disk, bulge, and gas velocities $V_k$, where $k$ refers to a component, are not their true velocities. Rather, they represent the velocities required to generate enough centrifugal acceleration to counter the gravitational acceleration produced by these components at a radial distance $r$, i.e., $\frac{V_k^2}{r} = - \frac{d\Phi_k}{dr}$. The gravitational potentials $\Phi_k$ are obtained by solving the Poisson equations for their corresponding densities such that they adhere to the observed surface brightness at $3.6 \mu m$ for the disk and the bulge. For gas, the HI/H$\alpha$ density profile is used with proper scaling ($1.33$) to incorporate the helium contribution. Thus, the total circular velocity $V_\text{tot}$ can be written as 
\begin{equation}
    V_\text{tot}^2 = V_\text{DM}^2 + \Upsilon_\text{disk} V_\text{disk} \vert V_\text{disk} \vert + \Upsilon_\text{bul} V_\text{bul} \vert V_\text{bul} \vert + V_\text{gas} \vert V_\text{gas} \vert. \label{eqn_vtot}
\end{equation}
Here, $V_\text{DM}$, calculated using Eq. (\ref{eqn:rotvel}), stands for the velocity component of the dark matter which forms an Einstein cluster. $V_\text{disk}$, $V_\text{bul}$, and $V_\text{gas}$ are the velocity components of the baryonic counterpart corresponding to the disc, bulge, and gas, respectively. The baryonic contributions must be scaled properly using the mass-to-light ratios for the disc and bulge components denoted as $\Upsilon_\text{disk}$, $\Upsilon_\text{bul}$ respectively. Note that the baryonic velocity components can be negative since they represent their gravitational accelerations. Therefore, Eq. (\ref{eqn_vtot}) is written in such a way as to preserve the proper sign of each component.

From the SPARC database, we use the observed velocity $V_\text{obs}$ along with its uncertainty $\delta V_\text{obs}$ and compare it with $V_\text{tot}$ to find the best-fit parameters for the Einstein cluster model. We use the Bayesian inference library $\textit{Bilby}$ \cite{bilby1, bilby2} along with the dynamic nested sampler $\textit{Dynesty}$ \cite{dynesty} to estimate the parameters in this model. Parameters related to the Einasto profile are $\rho_0$, $h$, $n$, and $R$, whereas Eq. (\ref{eqn_vtot}) suggests two more from the baryonic sector, i.e., $\Upsilon_\text{disk}$ and $\Upsilon_\text{bul}$. There is a general relation between the mass-to-light ratios from the stellar population synthesis models \cite{schombert_mcgaugh_2014} given as $\Upsilon_\text{bul} = 1.4 \Upsilon_\text{disk}$. However, following \cite{Lelli_2017} and \cite{Li_2019}, we use lognormal prior distributions on $\Upsilon_\text{disk}$ with mean value $0.5 M_\odot/ L_\odot$ and $\Upsilon_\text{bul}$ with mean $0.7 M_\odot/ L_\odot$ along with standard deviation $0.1$ dex. 
In addition, there are two more parameters tabulated in the SPARC database related to the baryonic part, i.e., distance $D$ and inclination $i$, rendering a total of $8$ parameters to be estimated for each galaxy. We use Gaussian prior distributions for the latter parameters with mean values set to be their tabulated values and standard deviations as given by their errors.

Any uncertainty in the measurement of $D$ leads to an adjusted value $D'$, directly leading to an adjustment of the halo radius as $R' = R (D'/D)$. Moreover, the baryonic velocity components are also scaled as $V_\text{k}' = V_\text{k} (\sqrt{D'/D})$, where k stands for disk, bulge, and gas. On the other hand, the observed velocity is measured independently of the radial distance. Therefore, $V_\text{obs}$ and $\delta V_\text{obs}$ remain unchanged for distance adjustment. However, the latter two are transformed due to an adjustment in inclination $i \to i'$ as $V_\text{obs}' = V_\text{obs} \sin{(i)}/\sin{(i')}$ and $\delta V_\text{obs}' = \delta V_\text{obs} \sin{(i)}/\sin{(i')}$. It is important to mention here that the inclination is assumed to be the same throughout a single galaxy.

\begin{figure}[]
\centering
\includegraphics[width=1\textwidth]{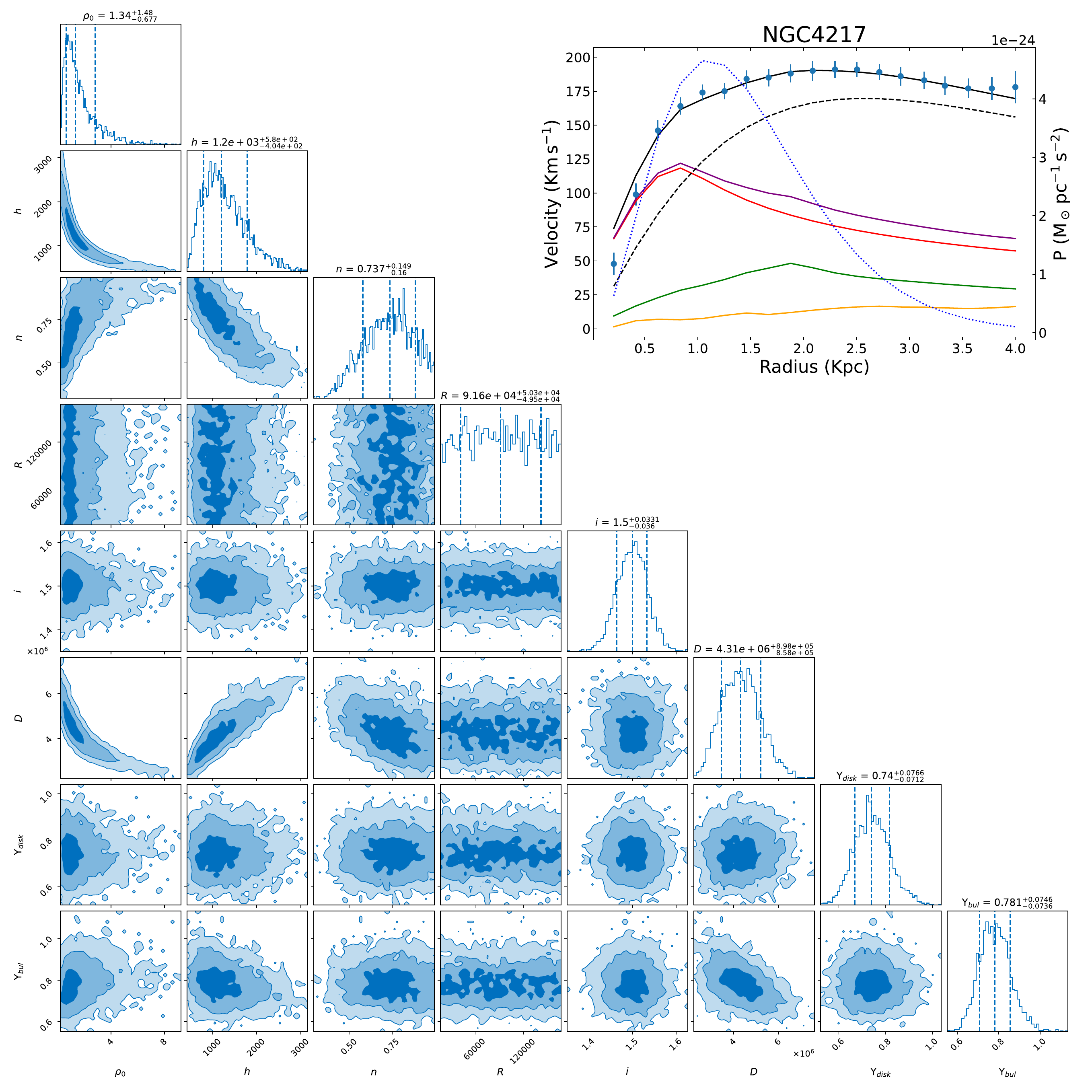}
\caption{Posterior distributions of $8$ model parameters are shown for a representative galaxy NGC4217. The dotted vertical lines in the 1D histograms denote 0.16, 0.5 (median), and 0.84 quantiles. \textbf{Top right panel:} The fitted rotation curve using the median values of the posterior distributions is shown in the black solid curve. Blue dots, along with the error bars, are the observed data. The baryonic components, scaled with proper mass-to-light ratios, are shown in colored solid curves. Orange indicates gas, green denotes disk, red stands for bulge, and violet is the total baryonic contribution. The black dashed curve suggests the contribution from the Einstein cluster. The blue dotted curve is the tangential pressure profile of the Einstein cluster. \label{fig:cornerplot}}
\end{figure}

For the remaining parameters, we set flat prior distributions. We consider the cluster radius $R$ to be larger than the largest observed radius of the galaxy in consideration. The posterior distributions for a representative galaxy NGC4217 are shown in Fig. (\ref{fig:cornerplot}). Note that, unlike the Newtonian Einasto halo models used in \cite{Li_2020}, $R$ is an additional parameter that directly determines the metric functions (see Eq. (\ref{eqn:gtt}) and (\ref{eqn:phi})). However, the posterior distribution for $R$ is found to be flat, which means that there is practically no dependence on $R$ at the galactic scale we are dealing with. In fact, we find that for all SPARC galaxies, $R$ follows a flat posterior distribution. Moreover, $i$, $D$, $\Upsilon_\text{disk}$ and $\Upsilon_\text{bul}$ deviate slightly from their initial standard values. From Eq. (\ref{eqn: Mass profile(r)}), we note that for sufficiently large $R$, $\rho_0 h^3$ approaches a constant value (considering $n$ fixed and $M$ saturates). This behaviour explains the parabolic type correlation (local degeneracy) we obtain for $\rho_0$ and $h$ posteriors.

\begin{figure}[]
\centering
\includegraphics[width=0.329\textwidth]{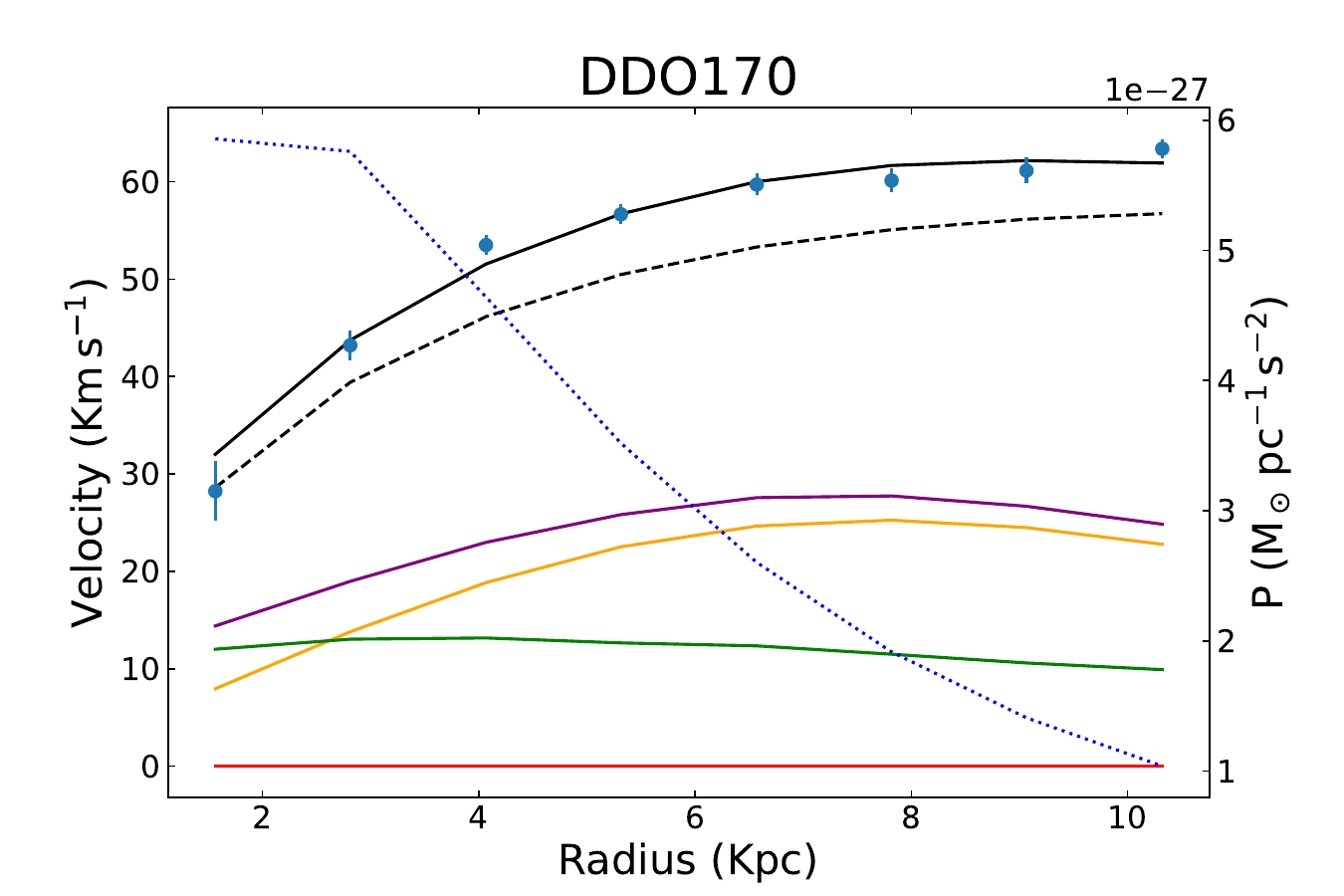}
\includegraphics[width=0.329\textwidth]{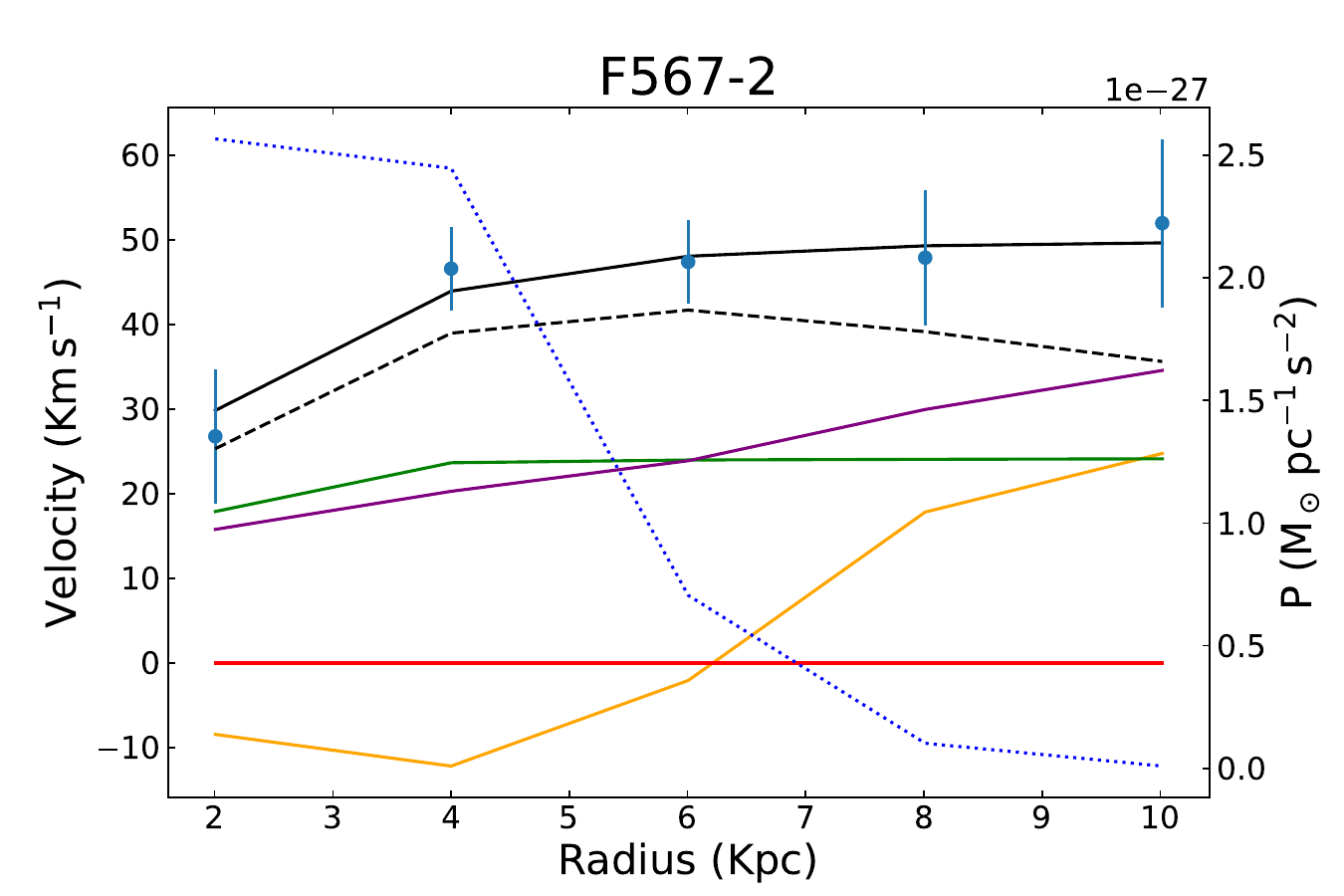}
\includegraphics[width=0.329\textwidth]{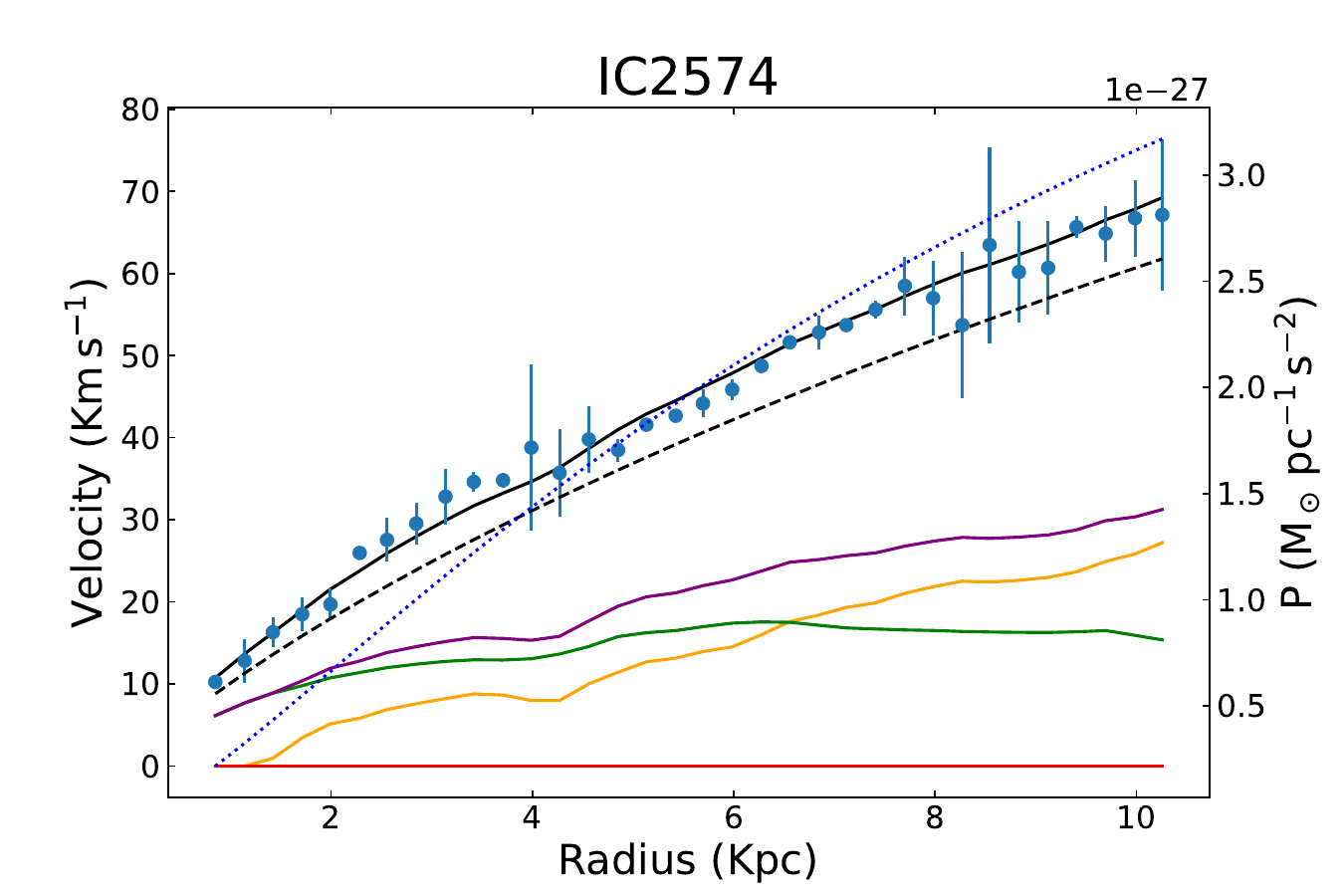}
\includegraphics[width=0.329\textwidth]{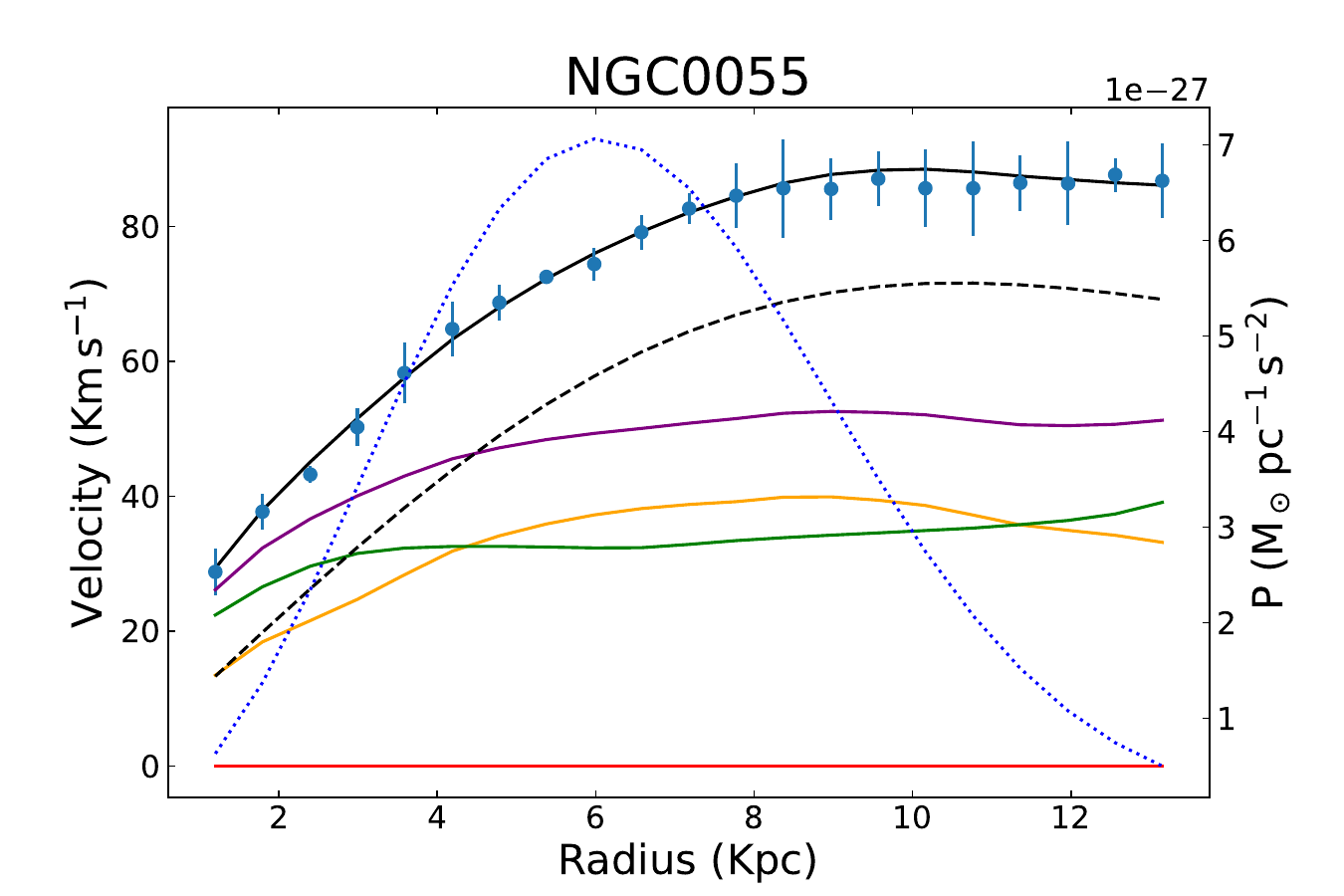}
\includegraphics[width=0.329\textwidth]{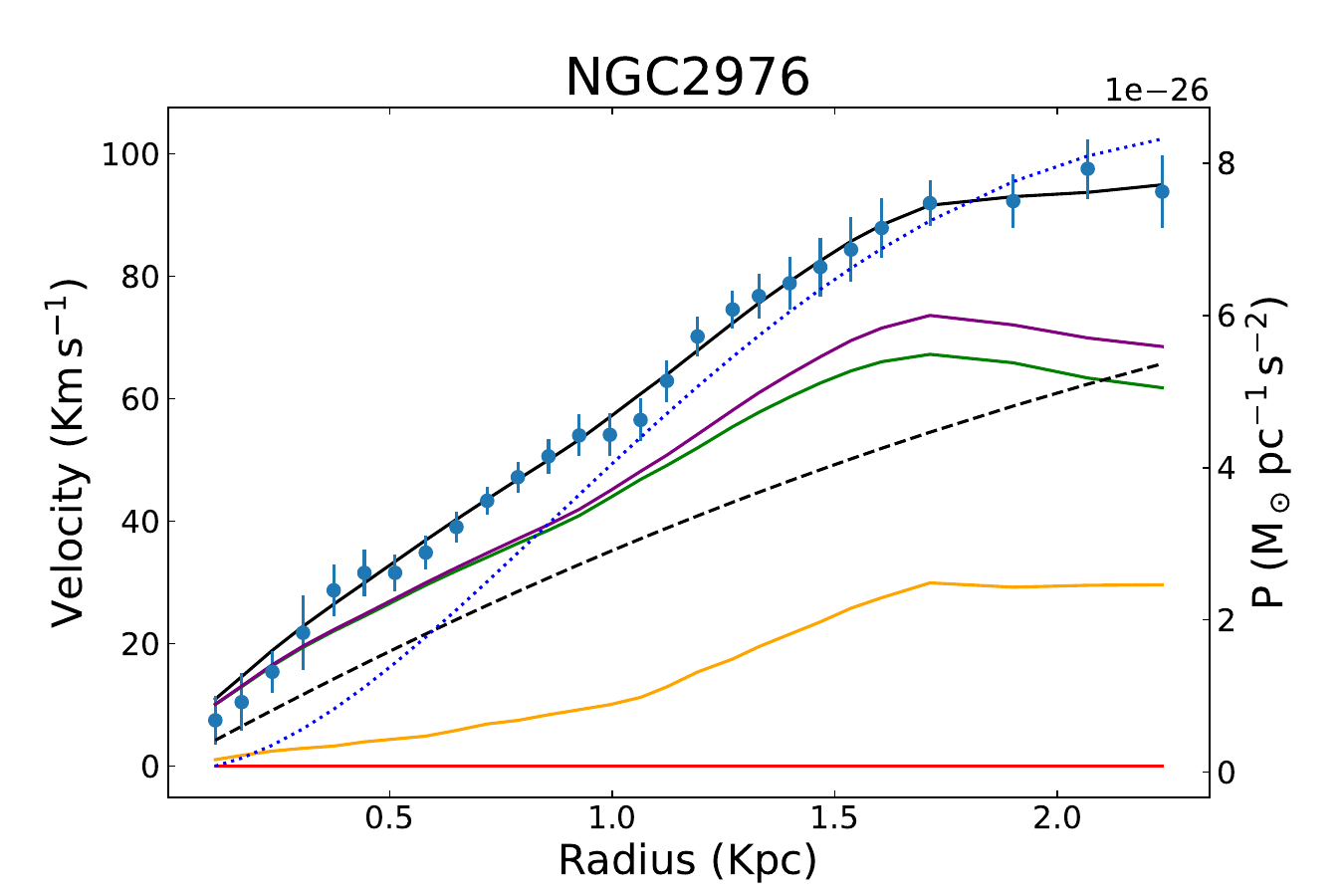}
\includegraphics[width=0.329\textwidth]{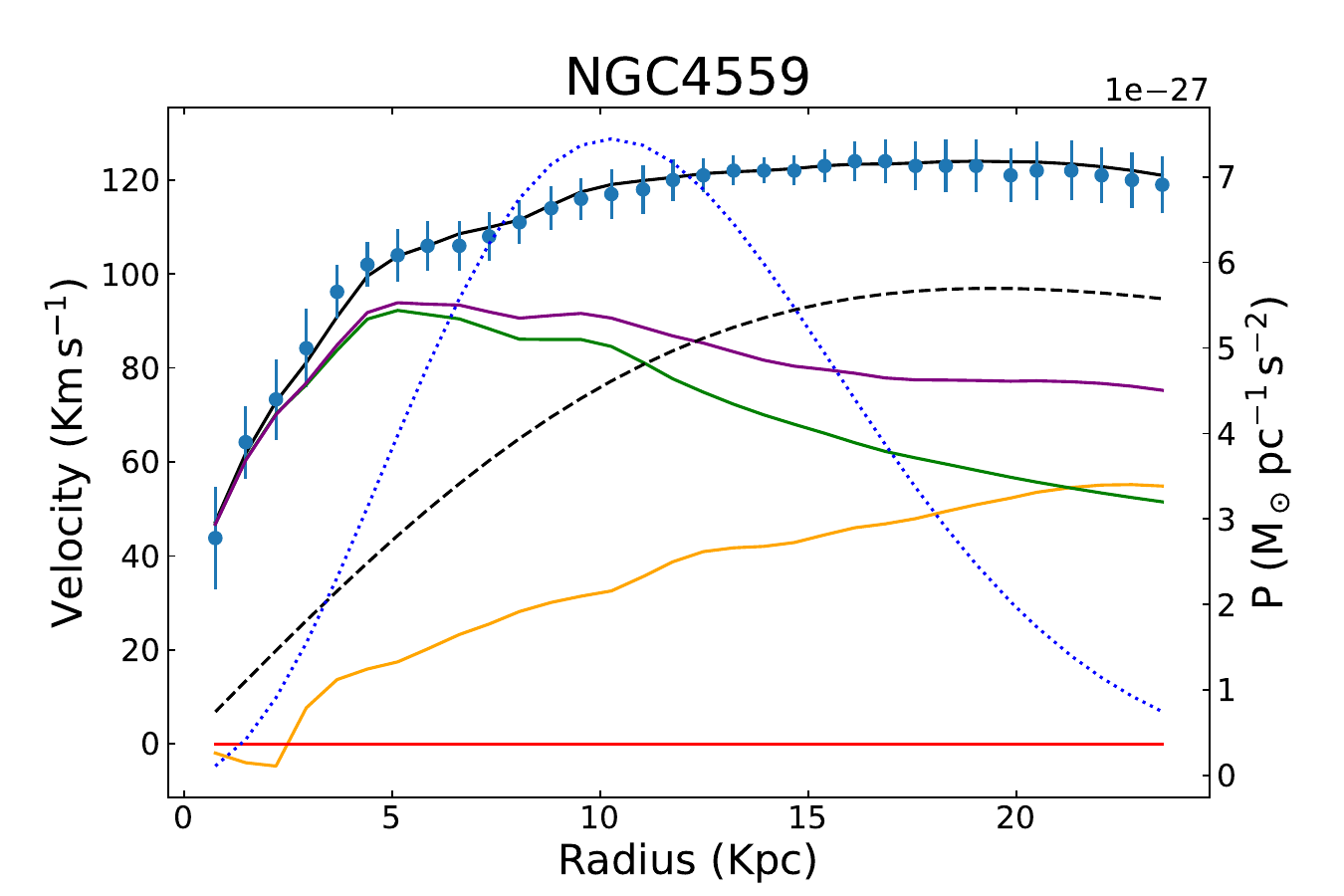}
\includegraphics[width=0.329\textwidth]{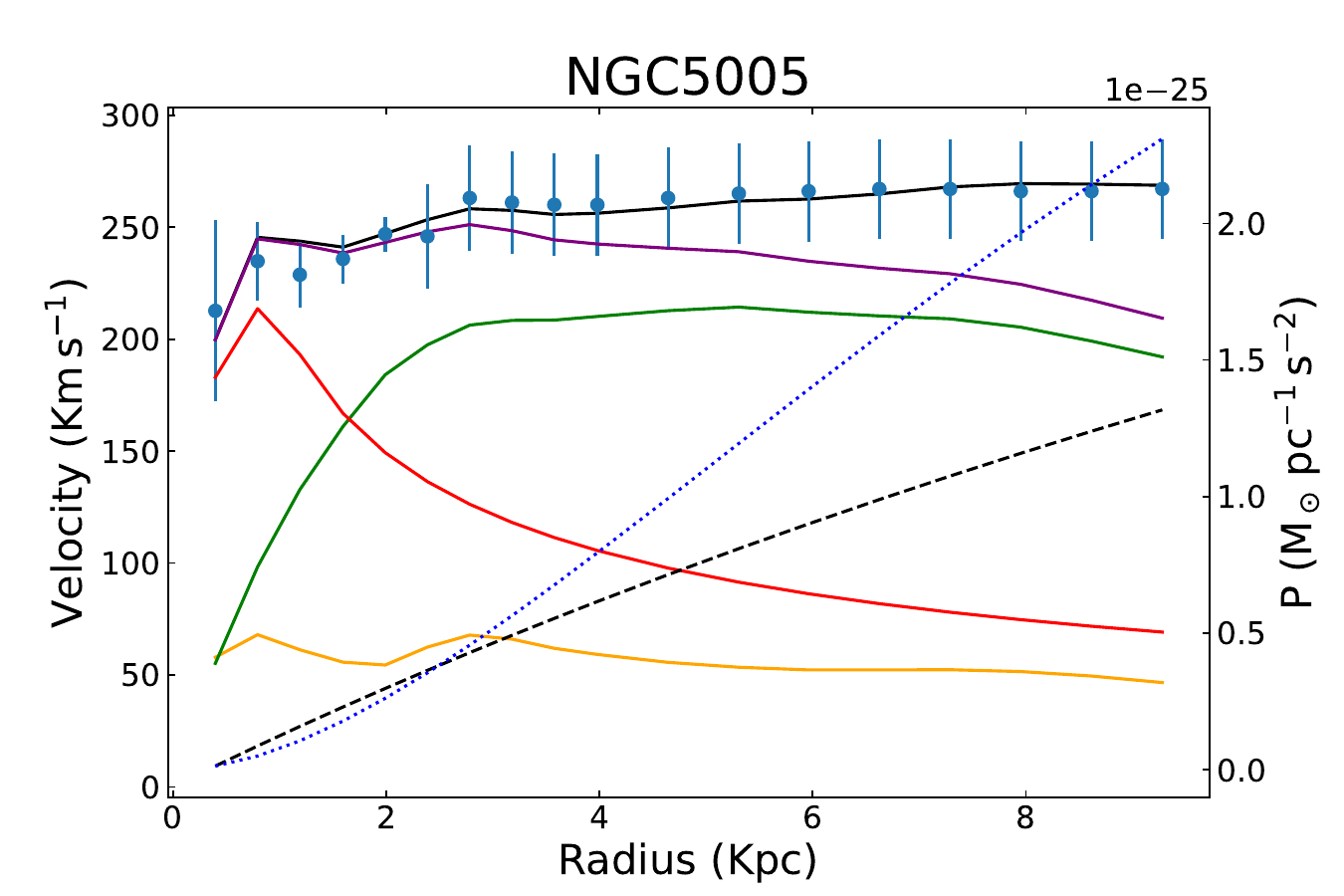}
\includegraphics[width=0.329\textwidth]{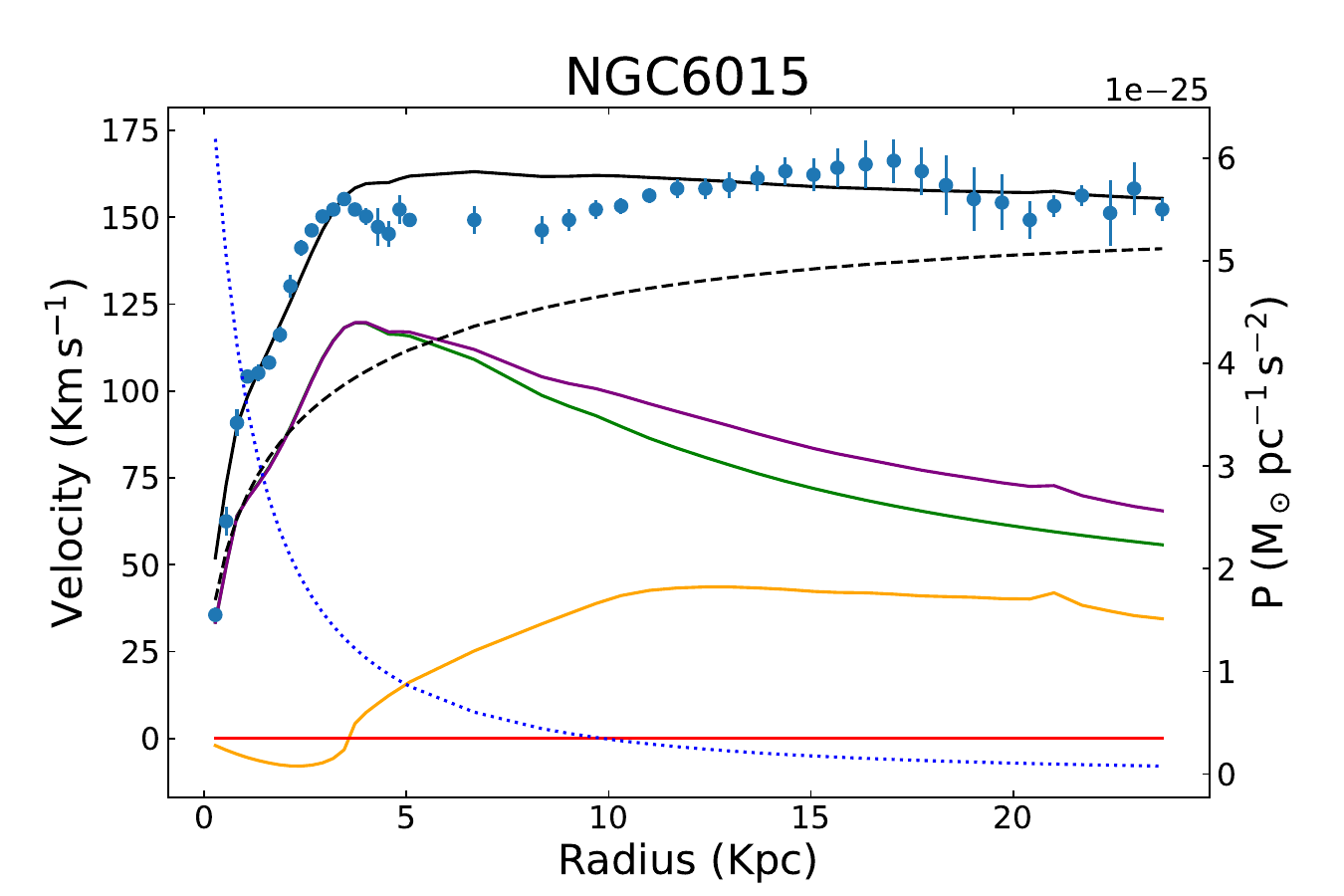}
\includegraphics[width=0.329\textwidth]{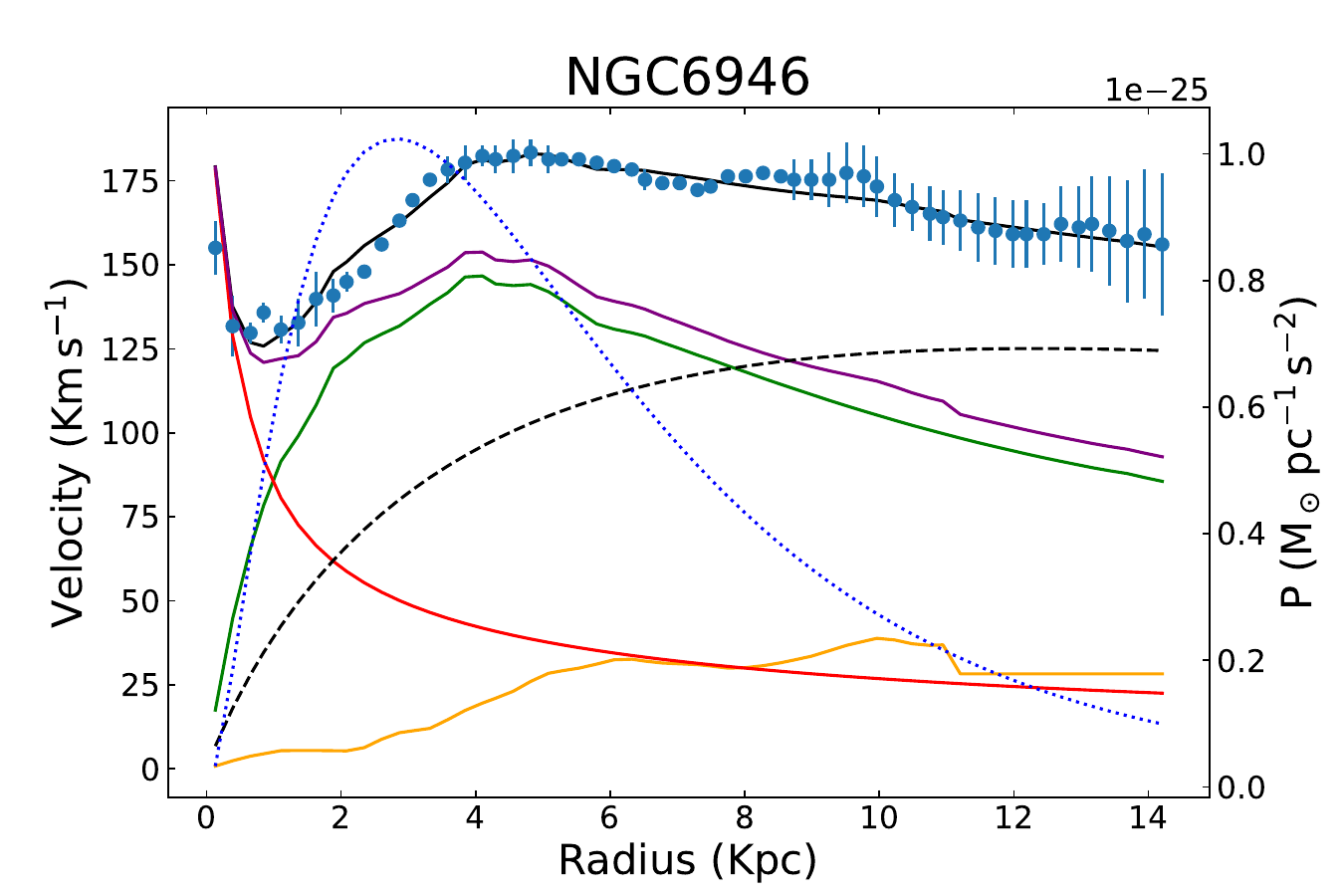}
\includegraphics[width=0.329\textwidth]{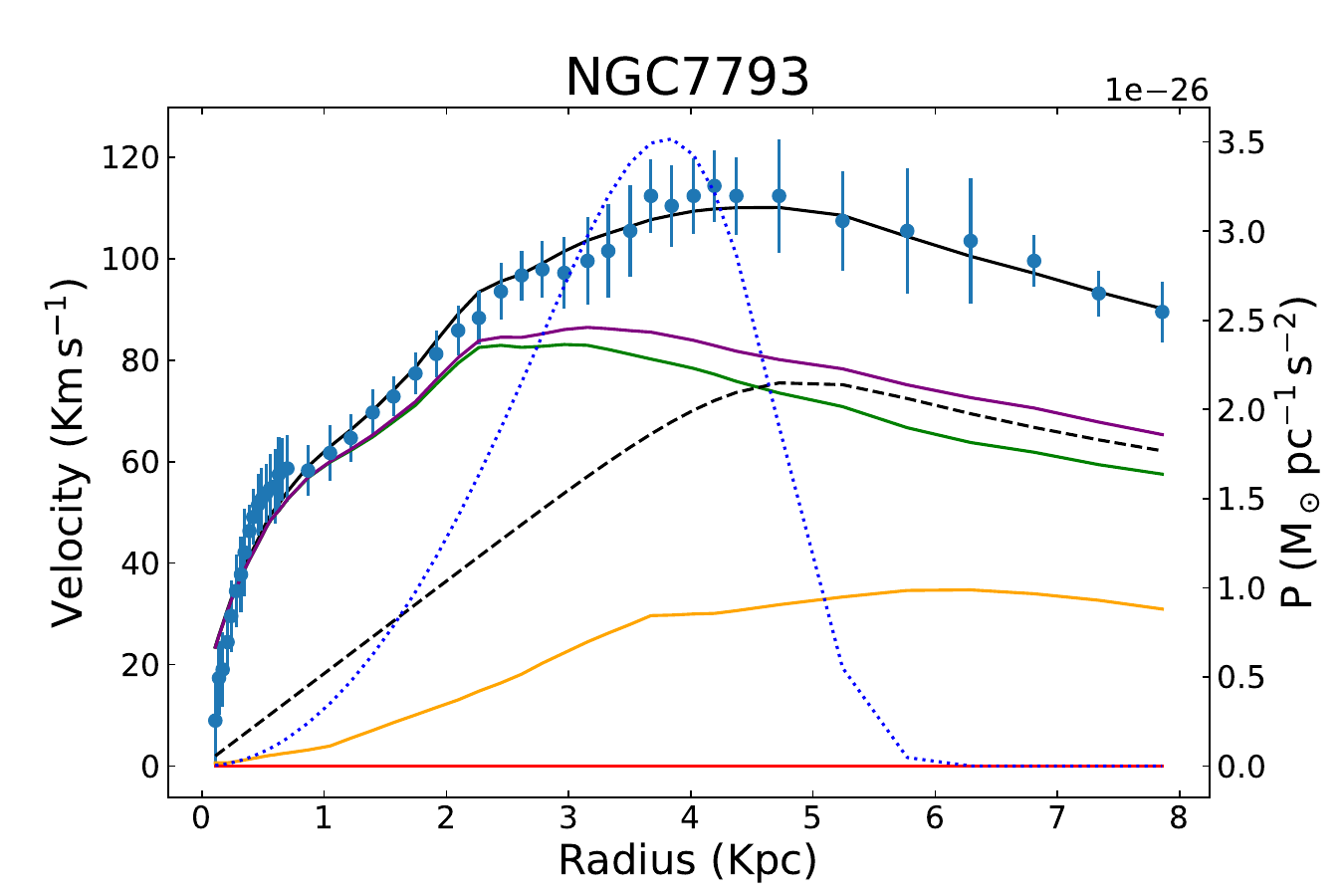}
\includegraphics[width=0.329\textwidth]{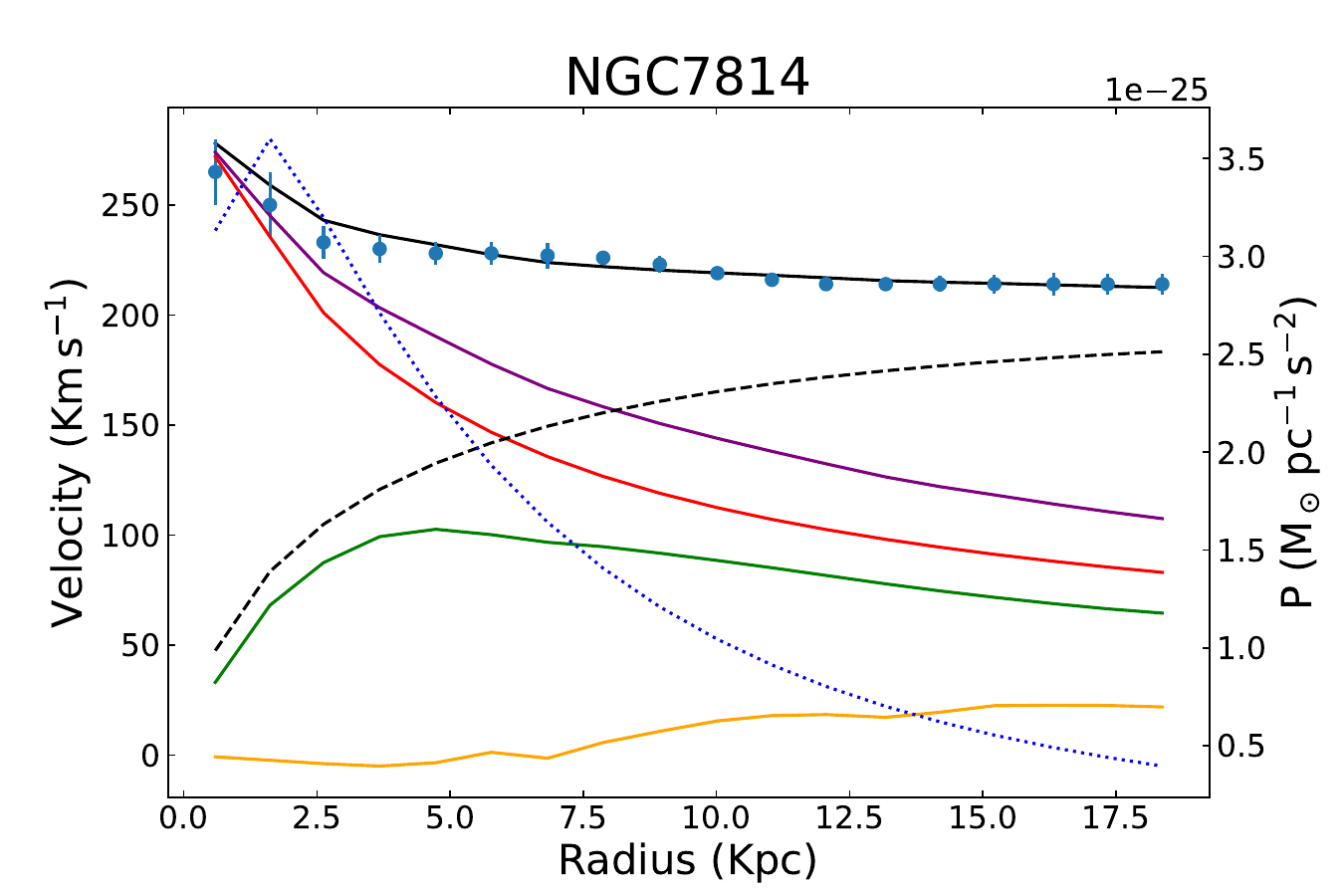}
\includegraphics[width=0.329\textwidth]{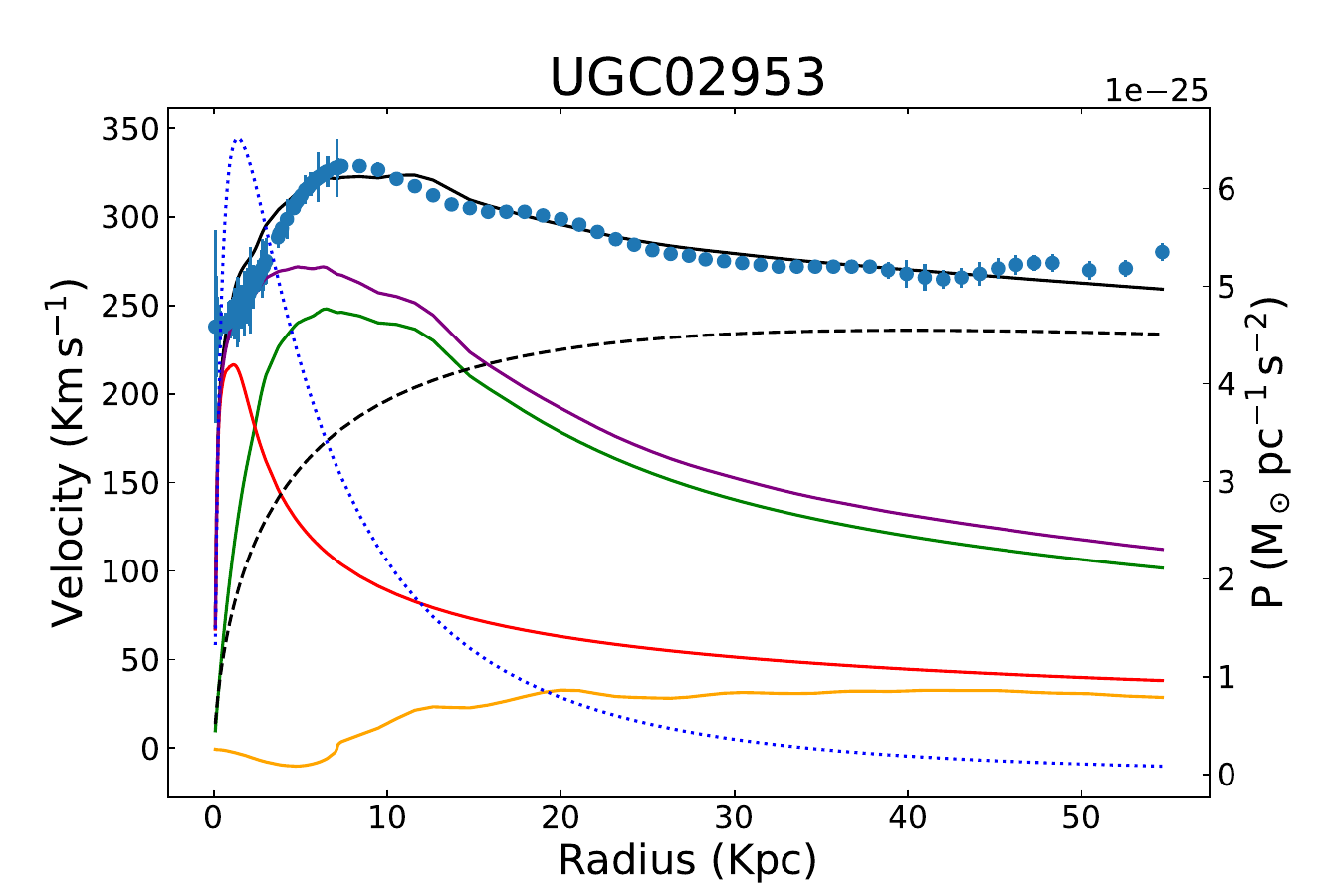}
\includegraphics[width=0.329\textwidth]{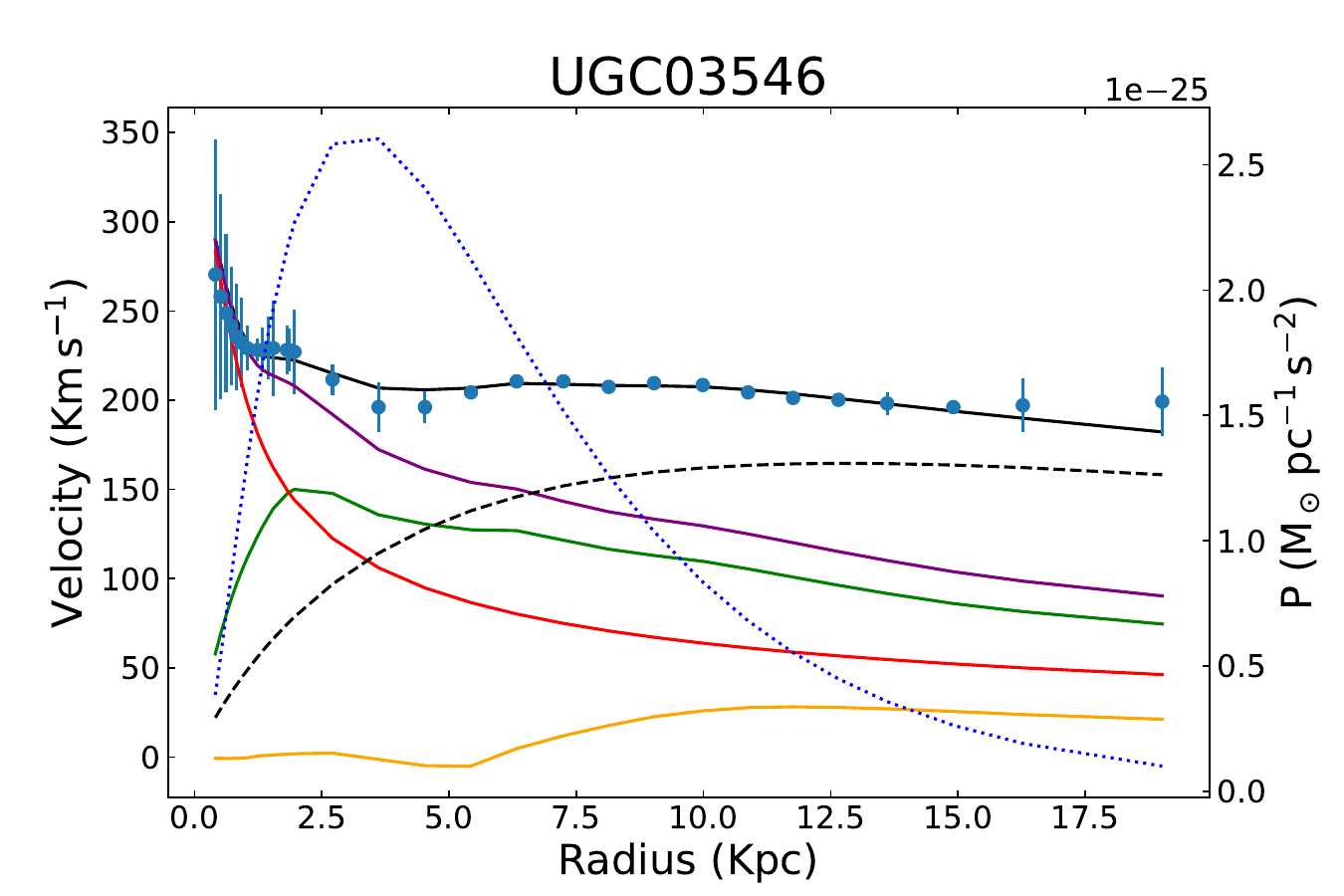}
\includegraphics[width=0.329\textwidth]{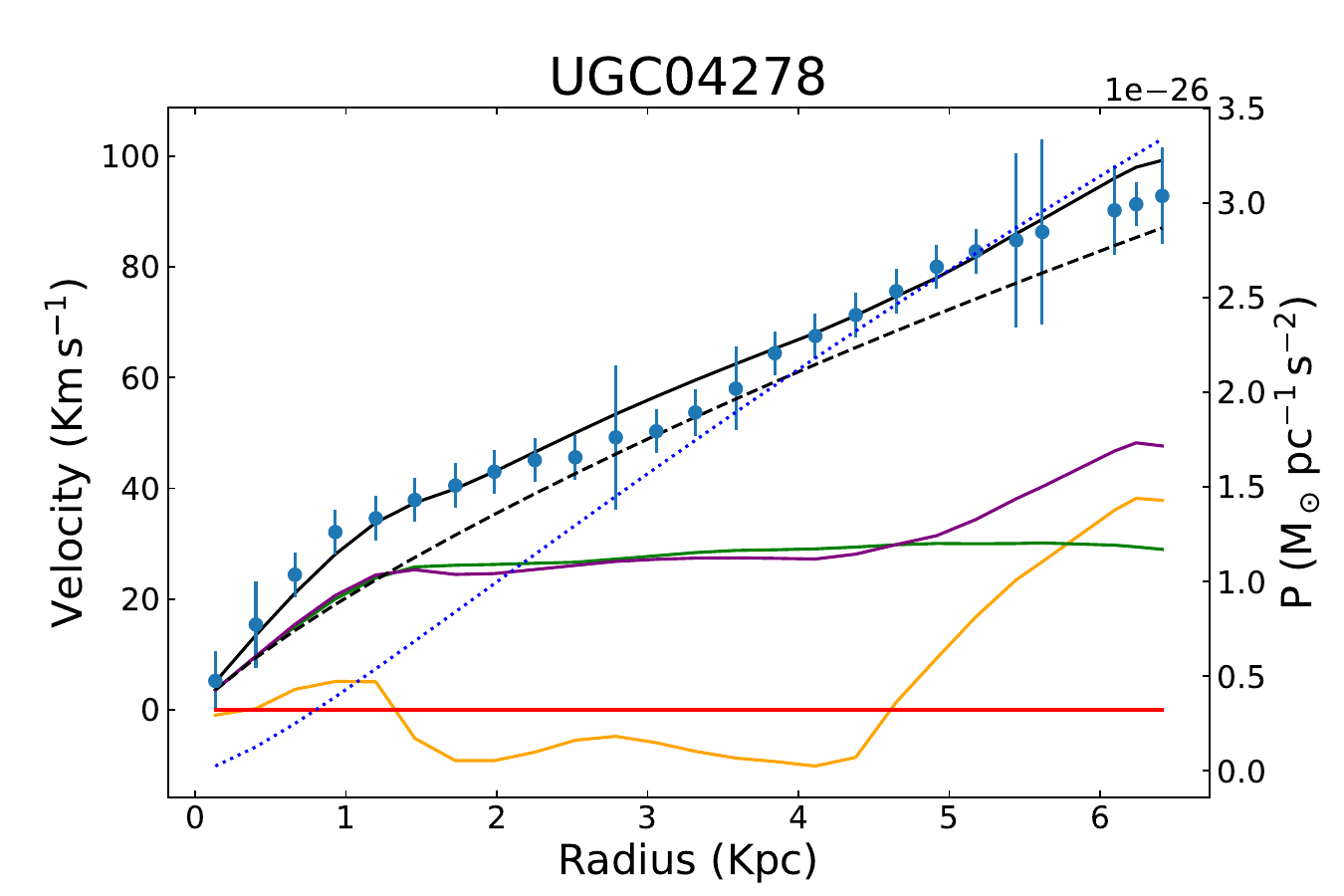}
\includegraphics[width=0.329\textwidth]{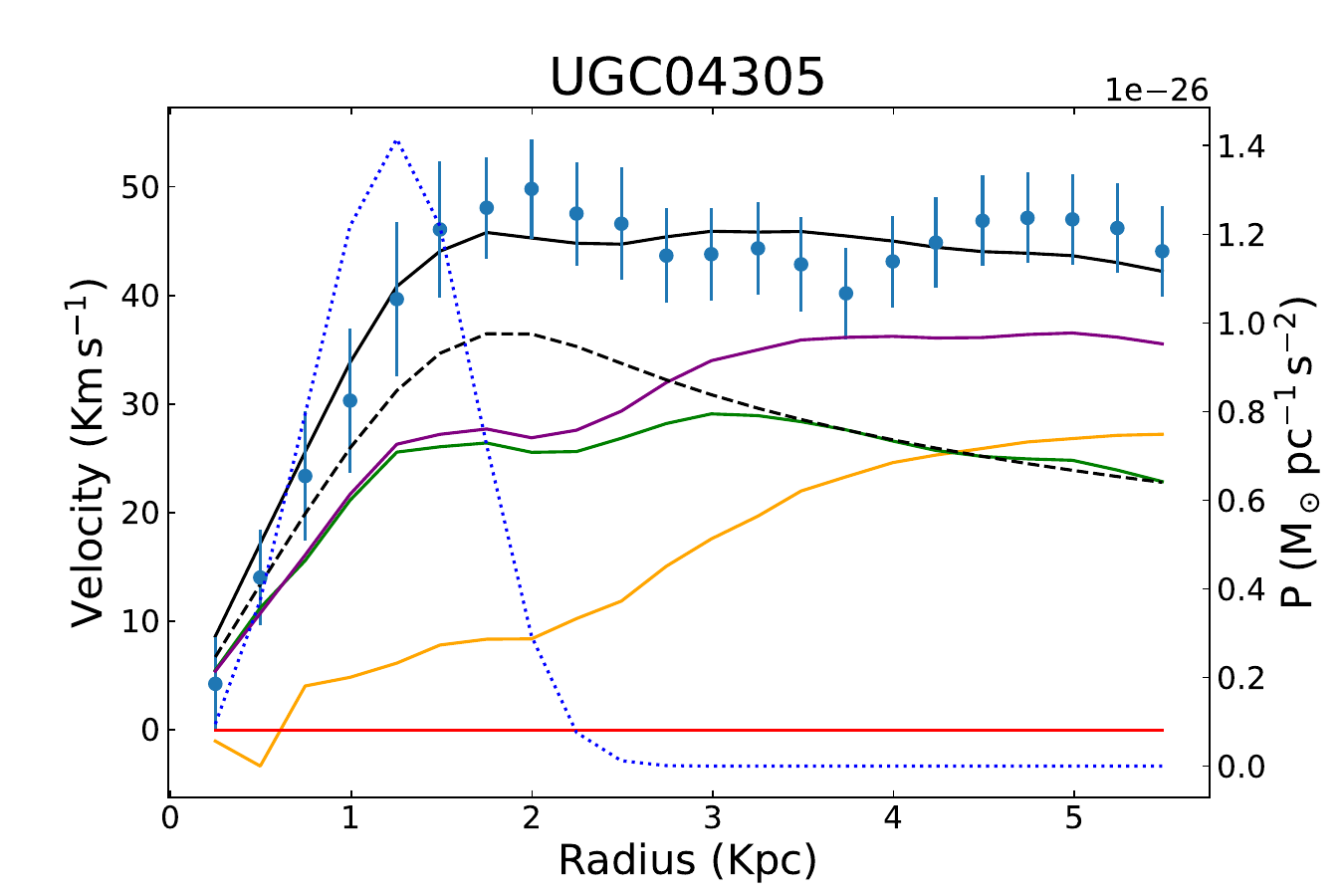}

\caption{Fitted rotation curves are shown for $15$ representative SPARC galaxies using the best-fit median of the model parameters. Details are the same as Fig. (\ref{fig:cornerplot}).    \label{fig:rotcurve}}
\end{figure}

Using the median values of the posteriors, we fit the observed rotation curve of NGC4217 with the Einstein cluster model (see the top right panel in Fig. (\ref{fig:cornerplot})). As shown in the figure, baryonic contributions to the rotation curve, including proper scaling of the mass-to-light ratios, are plotted separately. For this example, all three components, i.e., gas (shown in orange), disk (green), and the bulge (red), have non-zero contributions. As expected, their total baryonic contribution (shown in purple) cannot explain the observed velocity data (blue dots). When we add the contribution of the Einstein cluster on top of the baryonic contribution (using Eq. (\ref{eqn_vtot})), we get the best fit (denoted by the black curve). 

In Fig. (\ref{fig:rotcurve}), we also show the fitted curves for $14$ other representative galaxies. As clearly evident from the figure, the mass gap is filled by the contribution of the Einstein cluster. Our fits closely match \cite{Li_2020}, who used a Newtonian model for their analysis. It demonstrates the fact that SPARC galaxies demand the Newtonian limit of the general relativistic Einstein cluster model. However, unlike \cite{Li_2020}, we have nonzero tangential pressure inside the cluster. The pressure profile is also depicted in Fig. (\ref{fig:rotcurve}). We see different types of profiles depending on the density profiles of the galaxies. For instance, in cases of DDO170, F567-2, NGC6015, and NGC7814, the pressure profile starts from a higher value and gradually decreases towards the visible edges of these galaxies. On the other hand, for IC2574, NGC2976, NGC5005, and UGC04278, the tangential pressure increases from the galactic center towards the edge. For the rest, pressure first increases, attains a maximum, and then decreases to zero near the edge.  

\begin{figure}[h]
\centering
\includegraphics[width=0.75\textwidth]{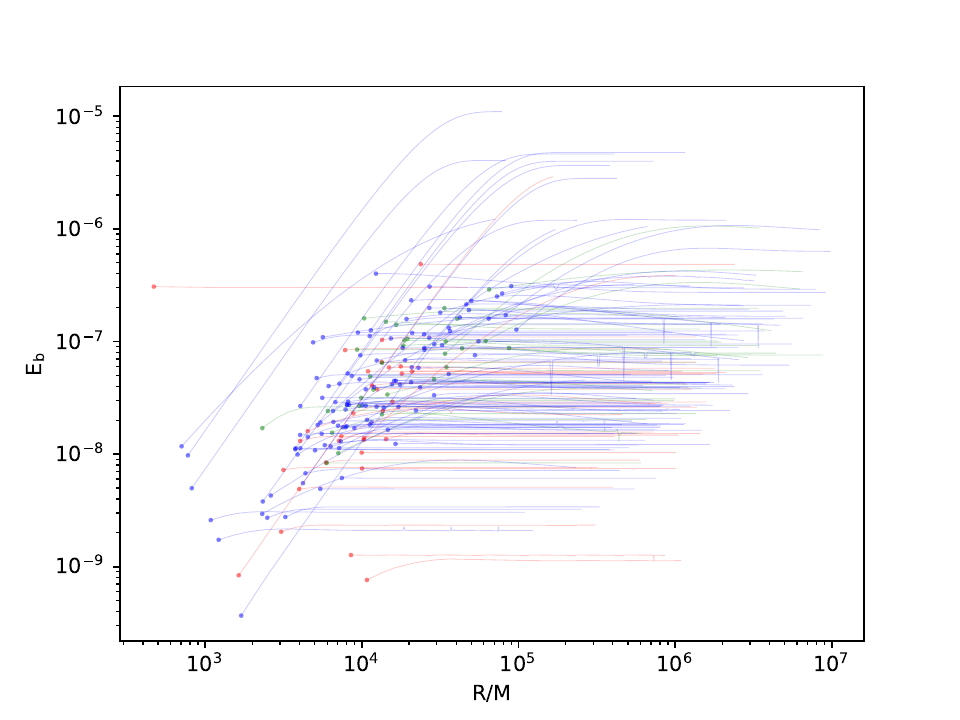}
\caption{Binding energy $E_b$ is plotted against $R/M$ for all SPARC galaxies. Each dot represents a galaxy. $E_b > 0$ confirms the stability of the Einstein clusters of all the galaxies. Green indicates NGCs, blue denotes UGCs and red represents other galaxies. A curve shows how binding energy varies if we increase the Einstein cluster radius up to $10$ times its starting value (denoted by a dot). \label{fig:EbRbplot}}
\end{figure}

In Fig. (\ref{fig:EbRbplot}), binding energies defined by Eq. (\ref{eqn:fractional binding energy}) of the galaxies are shown to be positive, which suggests that these halos modelled by Einstein clusters are stable. By varying the halo radius for each galaxy within its posterior range (same as the chosen prior range), we find that the binding energy remains positive. It should be noted here that the binding energy in Fig. (\ref{fig:EbRbplot}) is plotted considering the real units of $R$ and $h$, unlike Fig. (\ref{fig:Eb}) where these parameters are in units of $M$.  

\section{Conclusion}

In this work, we construct an Einstein cluster using the Einasto density profile, which is a popular and effective model for describing dark matter halos. Einstein cluster is described as a collection of collisionless particles that move in circular orbits in all possible directions and orbital radii. This is a general relativistic construction of an anisotropic system with zero radial but non-zero tangential pressure. In this case, the self-gravitational force is balanced by the centrifugal force acting on an orbiting particle. Our results are mentioned below, pointwise.
\begin{itemize}
    \item We revisit the formalism of the Einstein cluster, incorporate the Einasto density profile and study the stability of the cluster in two approaches. Firstly, using the effective potential, we obtain that the radius of the cluster must be $R \geq 6 M$ for a stable system. Secondly, we realise from the binding energy analysis that the total rest mass of all the particles must be higher than the gravitational mass of a stable cluster.
    \item Further, we investigate the validity of this model as a dark matter halo. We find the best-fit estimates of the model parameters using the SPARC galactic rotation curve data. Our fits adhere well with \cite{Li_2020}. The Einstein cluster model consists of an extra parameter $R$, which is absent in the Newtonian models. However, $R$ follows a flat posterior in the galactic scale limit.
    \item As an additional property, we quantitatively study the tangential pressure profile inside the Einstein clusters corresponding to the SPARC galactic halos. Since the galactic scale falls in the Newtonian regime and the rotational velocity is Keplerian, the magnitude of tangential pressure is very small compared to the energy density. However, it is an essential property for the existence of an Einstein cluster as a general relativistic model of galactic dark matter halo. We have further shown that the stability conditions are satisfied for the SPARC Einstein clusters.
    \item  Incorporating the Einasto density profile with the Einstein cluster model is analytically challenging due to the complex nature of the mass function; therefore, we have chosen the numerical approach to find the metric functions and the velocity profiles inside the cluster. 
\end{itemize}  

As a future project, we would like to investigate the stability of the cluster by analyzing gravitational perturbations in full detail. Analysing the effect of gravitational lensing in this model can be a good observational probe to test it and compare it with other standard dark matter models \cite{Bohmer_2007}. It may also be worthwhile to include a nonzero cosmological constant in our defining equations in order to make the model perhaps more realistic and noteworthy. 

\section{Acknowledgement} 
PB acknowledges financial support from the Science and Engineering Research Board, Government of India, File Number PDF/2022/000332.

\printbibliography

\end{document}